\documentclass[
 reprint,
superscriptaddress,
amsmath,amssymb,
prl
]{revtex4-2}

\usepackage{graphicx}
\usepackage{dcolumn}
\usepackage{bm}
\usepackage[mathlines]{lineno}

\usepackage{mathtools}
\usepackage{amsthm}

\usepackage{multirow}
\usepackage{array}
\usepackage{makecell}

\usepackage[colorlinks]{hyperref}
\hypersetup{linkcolor=magenta,urlcolor=blue,citecolor=blue,pdfstartview={FitH},hyperfootnotes=false,unicode=true}

\def\be{\begin{equation}}
\def\ee{\end{equation}}
\def\bea{\begin{eqnarray}}
\def\eea{\end{eqnarray}}
\def\nn{\nonumber}

\def\d{{\rm d}}

\def\Id{{\openone}}

\newcommand{\ket}[1]{\left|#1\right\rangle}
\newcommand{\bra}[1]{\left\langle #1\right|}
\newcommand{\braket}[1]{\left\langle #1 \right\rangle}

\newcommand{\abs}[1]{{\left| #1 \right|}}

\begin{document}

\title{Quantum Annealing Algorithms for Estimating Ising Partition Functions}

\author{Haowei Li}
\affiliation{School of Physics Science and Engineering, Tongji University, Shanghai 200092, China}
\affiliation{Laboratory of Quantum Information, University of Science and Technology of China, Hefei 230026, China}
\affiliation{Institute for Advanced Study, Tsinghua University, Beijing 100084, China}

\author{Zhiyuan Yao}
\email{yaozy@lzu.edu.cn}
\affiliation{Lanzhou Center for Theoretical Physics, Key Laboratory of Theoretical Physics of Gansu Province, Key Laboratory of Quantum Theory and Applications of MoE, Gansu Provincial Research Center for Basic Disciplines of Quantum Physics, Lanzhou University, Lanzhou 730000, China}

\author{Xingze Qiu}
\email{xingze@tongji.edu.cn}
\affiliation{School of Physics Science and Engineering, Tongji University, Shanghai 200092, China}

\date{\today}

\begin{abstract}
Estimating partition functions of Ising spin glasses is a cornerstone of statistical physics and computational science, yet it remains classically challenging due to its $\#$P-hard complexity. While Jarzynski's equality offers a theoretical pathway, its practical application is crippled at low temperatures by rare, divergent statistical fluctuations. Here, we introduce a quantum protocol that overcomes this fundamental limitation by synergizing reverse quantum annealing with optimized nonequilibrium initial distributions. Our method dramatically suppresses the estimator variance, achieving saturation in the low-temperature regime where existing methods fail. Numerical benchmarks on the Sherrington-Kirkpatrick spin glass and the 3-SAT problem demonstrate that our protocol reduces computational scaling exponents by over an order of magnitude (e.g., from $\sim 8.5$ to $\sim 0.5$), despite retaining exponential system-size dependence. Crucially, our protocol circumvents stringent adiabatic constraints, making it feasible for near-term quantum devices like superconducting qubits, trapped ions, and Rydberg atom arrays. This work provides a methodological framework for quantum-enhanced estimation in spin glass thermodynamics and beyond by harnessing non-adiabatic quantum dynamics to address a classically difficult problem.

\end{abstract}

\maketitle


{\it Introduction.---}
Ising spin glasses (ISGs) play a pivotal role in both fundamental research and practical applications across diverse disciplines \cite{Binder_1986_RMP, Nishimori_Book_SpinGlass,Stein_Book,Charbonneau_2023}, including statistical physics \cite{King_2023_Nature}, combinatorial optimization \cite{Kirkpatrick_1983_Science}, and machine learning \cite{Melko_2019_NP}.
Despite its broad relevance, a longstanding challenge persists: estimating the Ising partition functions (IPFs) of these complex systems \cite{Huang_2008}, which is classically difficult, being $\#$P-hard in the worst case \cite{Lidar_2004_NJP}.
Conventional approaches to this problem face fundamental roadblocks, particularly in the physically interesting low-temperature regime.

The Jarzynski's equality (JE) \cite{Jarzynski_1997_PRL, Tasaki_2000_arXiv, Kurchan_2001_arXiv, Mukamel_2003_PRL, Esposito_2009_RMP, Campisi_2011_RMP}, an elegant identity from nonequilibrium statistical mechanics, theoretically connects the free energy difference between two systems to the average work done during a transformation between them. However, its practical utility for IPF estimation is crippled at low temperatures by rare events dominating the exponential average \cite{Jarzynski_2006_PRE, Xiao_2014_PRE}.
Classical methods based on Markov-chain Monte Carlo (MCMC) \cite{Daniel_2009, Kolmogorov_2018} can exhibit extremely long equilibration and autocorrelation times in rugged low-temperature energy landscapes, resulting in trapping within metastable basins. While enhanced sampling methods offer partial mitigation \cite{Neal2001AIS, BergNeuhaus1992, WangLandau2001a, WangLandau2001b}, low-temperature barriers remain a persistent bottleneck in spin-glass regimes due to rapidly increasing mixing and round-trip times \cite{Dayal2004WLspinGlass}. 
In view of this fundamental barrier, advanced quantum algorithms have emerged as promising candidates for circumventing classical limitations. Recent proposals include quantum MCMC \cite{Montanaro_2015}, quantum phase estimation \cite{Poulin_2009_PRL}, and one clean
qubit (DQC1)-based algorithms  \cite{Chowdhury_2021_PRA, Jackson_2023_PRA}.
However, these approaches remain largely confined to idealized settings and demand quantum resources beyond what is currently feasible on noisy intermediate-scale quantum (NISQ) devices \cite{Bharti_2022_RMP}.

\begin{figure}[tp!]
\centering
\includegraphics[width=0.48\textwidth]{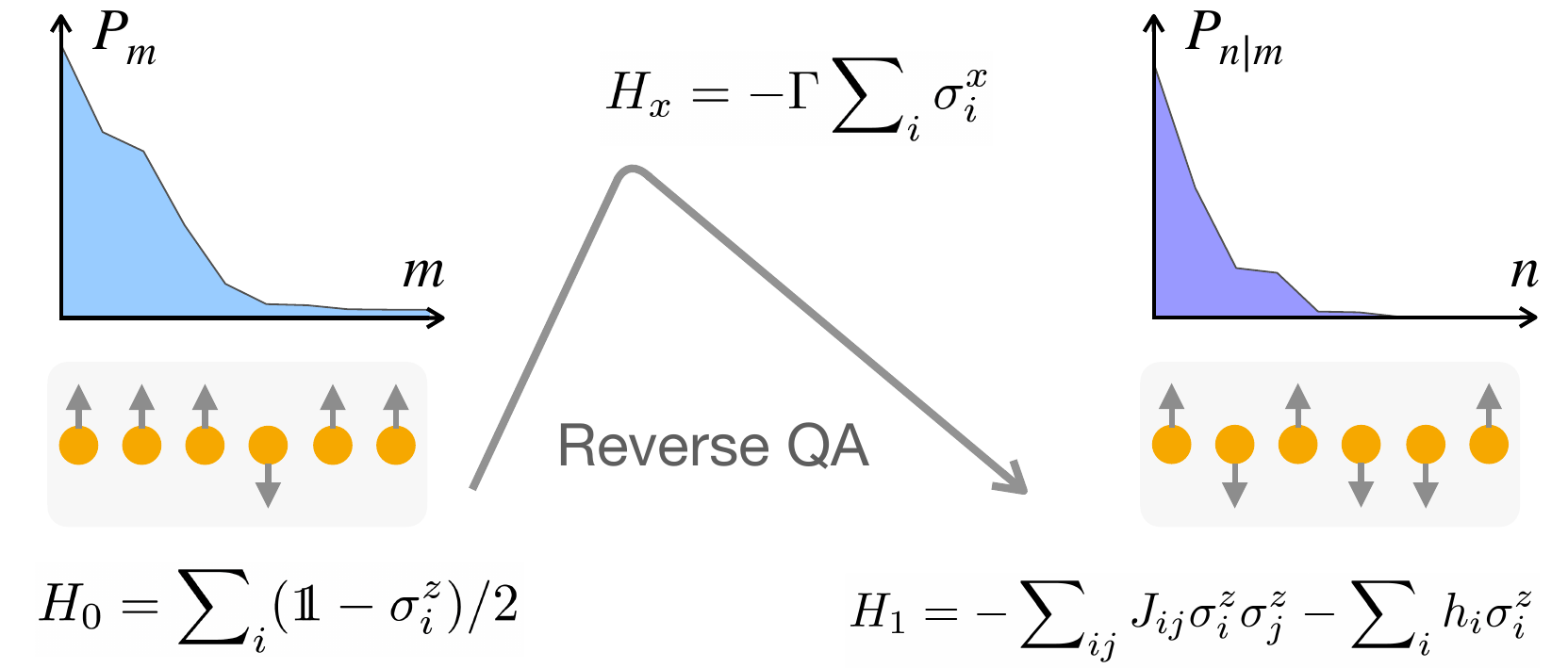}
\caption{
Schematics of our protocol for estimating IPFs.
The protocol begins by preparing the system in an initial state sampled from a distribution $P_m$. Unlike conventional JE approaches that assume thermal equilibrium (e.g., Gibbs distributions), $P_m$ is intentionally designed to deviate from equilibrium to suppress the variance.
The system then goes through a nonequilibrium RQA process [Eq.~\eqref{eq:H_RQA}], in which the Hamiltonian transitions from a trivial free model $H_0$ to the desired target ISG $H_1$ with the aid of a driver $H_x$.
Finally, projective measurements in the eigenbasis of $H_1$ yield outcomes governed by the conditional probability $P_{n|m}$ [see details above Eq.~\eqref{eq:Z_1}].
These outcomes are processed via the estimator to reconstruct the IPFs of $H_1$ [Eq.~\eqref{eq:Estimator}].
}
\label{fig:Fig_1}
\end{figure}

Here, we introduce a new algorithmic paradigm that is explicitly designed to overcome these barriers by harnessing the native physics of NISQ hardware. As illustrated in Fig.~\ref{fig:Fig_1}, our approach is a synergistic hybrid algorithm that combines a classically optimized, nonequilibrium initial state distribution with a quantum evolution generated by reverse quantum annealing (RQA) \cite{Alejandro_2011, Chancellor_2017, King_2018_Nature, Ohkuwa_2018_PRA, Yamashiro_2019_PRA}. The choice of RQA is not one of convenience but is fundamental to the protocol's performance and feasibility. First, RQA is a hardware-native physical process on leading quantum annealing platforms, minimizing control overhead. Second, RQA excels at the local refinement of known classical states, making it the ideal quantum engine to explore the promising regions of the solution space identified by our classical optimization. Third, and most profoundly, our protocol operates optimally in a non-adiabatic regime. An infinitely long, perfectly adiabatic evolution is not only unnecessary but suboptimal. This means that the primary constraint of NISQ hardware---limited coherence time---is not a fatal flaw for our protocol but is instead aligned with its optimal operating regime.
Our protocol can thus be implemented on current platforms like superconducting qubits \cite{Albash_2018_PRX, Hauke_2020_IOP, Miessen_2024_PRXQuantum}, trapped ions \cite{Guo_2024_Nature, Kihwan_2025_PRL}, and Rydberg atom arrays \cite{Qiu_2020_PRXQ, Ebadi_2022_Science}.

Our work demonstrates two key advances.
First, the protocol completely circumvents the rare-event bottleneck of JE-based approaches, achieving saturation of the estimator variance at low temperatures and reducing the computational scaling exponent by over an order of magnitude.
Second, our protocol positions quantum annealing (QA) as a versatile tool for estimating IPFs, which extends conventional QA's utility beyond the ground-state search of ISGs \cite{Lucas_2014_FP, Albash_2018_RMP, Crosson_2021_NRP}. This enriches the research that utilizes QA-enabled Gibbs sampling \cite{Wild_2021_PRL, Vuffray_2022_PRXQuantum, Shibukawa_2024_PRR} in approximating the finite-temperature properties of ISGs.
By exploiting quantum dynamics to sidestep the critical failures of existing methods, our protocol establishes a practical and powerful application for near-term quantum processors in tackling classically difficult problems. This work advances applications in ISG physics, optimization, and machine learning, where IPFs estimation is critical yet classically difficult.


{\it Reverse quantum annealing protocol.---}
RQA enables state-selective initialization of QA dynamics---a feature critical for applications such as hybrid quantum-classical optimization \cite{Chancellor_2017} and quantum simulations \cite{King_2018_Nature}.
The RQA protocol can be described by the following form of a time-dependent Hamiltonian \cite{Ohkuwa_2018_PRA, Yamashiro_2019_PRA}
\bea
H(t) = s(t)H_1 \, &+& \, [1-s(t)]\lambda(t) H_x  \nn \\
&+& \, [1-s(t)][1-\lambda(t)]H_0\, .
\label{eq:H_RQA}
\eea
Here, the time-dependent parameters $s(t)$ and $\lambda(t)$ both grow from $0$ to $1$ over the course of the unitary time evolution $U(\tau) = \mathcal{T}\exp[-i\int_{0}^{\tau} H(t)\,\d t]$ with $\tau$ the total time and $\mathcal{T}$ the time-ordering operator. The system thus transitions from the initial Hamiltonian $H_0$ to the target Hamiltonian $H_1$, aided by the driver Hamiltonian $H_{x}$. In our context of the ISG problem, $H_0=\sum_{i=1}^N  (\Id-\sigma^z_i)/2$, $H_1 = -\sum^N_{i,j=1} J_{ij}\sigma^z_i\sigma^z_j - \sum^N_{i=1} h_i \sigma^z_i$, and $H_x = -\Gamma\sum^N_{i=1}\sigma^x_i$ where $\Id$ is the identity operator, $\{\sigma^{x/z}_i\}$ are the Pauli-$x/z$ operators, and $N$ is the system size. The spin-spin couplings $J_{ij}$ and the local fields $h_i$ defines a model instance \cite{Lucas_2014_FP,  Albash_2018_RMP}, and $\Gamma$ sets the scale of quantum fluctuations for state transitions. Hereafter, we set $\hbar=1$ and $k_{\rm B} = 1$, use $\Gamma$ as our energy unit, and choose $s(t)=\lambda(t) = t/\tau$.
The choice of $H_0$ is physically motivated, as its eigenstates are computational basis states that can be efficiently prepared on a wide range of quantum hardware platforms, making RQA the natural framework for initiating the desired quantum evolution.

Let $\{E^0_m,\, \ket{\psi^0_m}\}$ and $\{E^1_n,\, \ket{\psi^1_n}\}$ denote the ascending eigenvalues and corresponding eigenstates of $H_0$ and $H_1$, respectively. The system is initialized by randomly sampling $\ket{\psi^0_m}$ according to a sampling function $P_m$, which satisfies $P_m > 0$ and $\sum^D_{m=1}P_m=1$ with $D=2^N$ the dimension of the Hilbert space.
Under the RQA dynamics, the initial state evolves to $\ket{\psi(\tau)} = U(\tau)\ket{\psi^0_m}$. The following projective measurement in the eigenbasis of $H_1$ yields a trajectory $\ket{\psi^0_m}\to \ket{\psi^1_n}$ with a conditional probability $P_{n|m}=\abs{\bra{\psi^1_n}U(\tau)\ket{\psi^0_m}}^2$. Due to the unitarity of $U(\tau)$, we have $\sum^D_{m=1}P_{n|m}=1$.
Now we are ready to reconstruct the IPFs of $H_1$ as
\bea
Z_1(\beta)
&=& \sum\nolimits^D_{n=1} \exp \left(-\beta E^1_n\right)  \nn \\
&=& \sum\nolimits^D_{m,n=1} \frac{\exp \left(-\beta E^1_n\right)}{P_m} P_{n|m} P_m\, ,
\label{eq:Z_1}
\eea
where $\beta=1/T$ is the inverse temperature.
This estimator leverages correlations between initial ($P_m$) and transition ($P_{n|m}$) probabilities to bypass direct summation over $D$ states.

\begin{figure*}[htp!]
\centering
\includegraphics[width=1\textwidth]{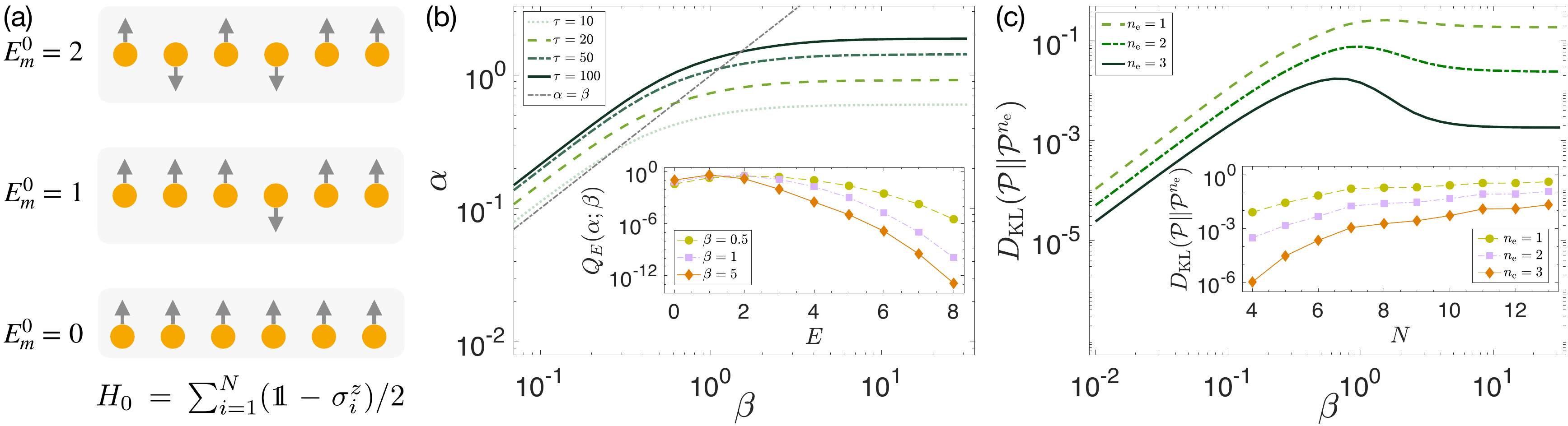}
\caption{
Approximation strategies for optimized sampling functions in the SK spin-glass model.
(a) Initial spin configurations.
Arrows represent spins in the eigenstates of $\sigma^z$, $\{\ket{\uparrow},\ket{\downarrow}\}$ (eigenvalues $+1$ and $-1$, respectively).
The initial product state $\ket{\psi^0_m}=\otimes^N_{i=1}\ket{\gamma_i}$ ($\ket{\gamma_i}\in\{\ket{\uparrow},\ket{\downarrow}\}$), is an eigenstate of $H_0$.
The number of $\ket{\downarrow}$ in $\ket{\psi^0_m}$ equals its energy $E^0_m$, with $I_E=\{m|E^0_m=E\}$ indexing degenerate states of energy $E$.
(b) Gibbs approximation of $\mathcal{P}_m(\beta)$.
The optimized sampling function $\mathcal{P}_m(\beta)$ is approximated by a Gibbs distribution $P^G_m(\alpha)$ for total evolution times $\tau = 10,20,50,100$.
The gray dashed-dotted $\alpha=\beta$ line is drawn for reference purposes.
The inset shows the distribution $Q_E(\alpha;\beta)$ for $\beta=0.5,1,5$ with $\tau=100$.
Here, we have chosen the system size $N=8$.
(c) Perturbative approximation of $\mathcal{P}_m(\beta)$.
The Kullback--Leibler divergence $D_{\rm KL}(\mathcal{P}\|\mathcal{P}^{n_{\rm e}})=\sum_m \mathcal{P}_m\log(\mathcal{P}_m/\mathcal{P}^{n_{\rm e}}_m)$ \cite{Kullback_1951} quantifies the difference between $\mathcal{P}_m(\beta)$ and $\mathcal{P}^{n_{\rm e}}_m(\beta)$ for $n_{\rm e} = 1,2,3$.
Main panel: $\beta$-dependence ($N=8$, $\tau=100$).
Inset: System-size scaling ($N=4\text{--}13$, $\tau=100$, $\beta=10$).
All data are averaged over $10^3$ Hamiltonian instances.
}
\label{fig:Fig_2}
\end{figure*}

Note that choosing $P_m= P^{\rm G}_m(\beta) \equiv \exp(-\beta E^0_m)/Z_0(\beta)$ (Gibbs distribution) will reduce Eq.~\eqref{eq:Z_1} to the celebrated JE: $Z_1(\beta)/Z_0(\beta) =\exp(-\beta \Delta F)=\braket{\exp(-\beta W)}$ \cite{Jarzynski_1997_PRL, Tasaki_2000_arXiv, Kurchan_2001_arXiv, Mukamel_2003_PRL}.
Here, $Z_0(\beta) = [1+\exp(-\beta)]^N$ is the trivial partition function of $H_0$, $\Delta F = -\beta^{-1}\log[Z_1(\beta)/Z_0(\beta)]$ is the free energy difference, and the expectation value $\braket{ \exp(-\beta W)}=\sum^D_{m,n=1}\exp(-\beta W_{nm})P_{n|m} P_m$ with $W_{nm}=E^1_n-E^0_m$ the quantum work defined through a two-point measurement scheme \cite{Tasaki_2000_arXiv, Kurchan_2001_arXiv, Mukamel_2003_PRL}.
By enabling efficient preparation of initial Gibbs ensembles, this protocol establishes quantum processors as practical tools for ISG thermodynamics---directly supporting applications such as validating fluctuation theorems in many-body regimes \cite{Hahn_2023_PRX} and estimating free-energy for disordered systems \cite{Bassman_2022_PRL}.

In the following, we focus on estimating the IPFs of $H_1$ based on Eq.~\eqref{eq:Z_1}, to achieve contrasted protocol advantages over conventional JE-based approaches.
To demonstrate the efficiency of our protocol, we benchmark it against two canonical ISG models: the Sherrington--Kirkpatrick (SK) spin glass \cite{Sherrington_1975_PRL,Kirkpatrick_1978ir} and the random 3-SAT \cite{Barthel_2002_PRL_3SAT}.
The Hamiltonian of the SK model reads
$
H_{\rm SK} = \frac{1}{\sqrt{N}}\sum^N_{(j\neq i)=1}J_{ij}\sigma^z_i\sigma^z_j + \sum^N_{ i=1}h_{i}\sigma^z_i\, ,
$
where $J_{ij}$ and $h_{i}$ are independent variables sampled from the standard normal distribution.
And the random 3-SAT is a fundamental Boolean satisfiability problem. The hard instances are generated via a physics-inspired protocol with planted solutions, ensuring controlled benchmarking in classically difficult regimes \cite{OurSM}.


{\it Optimized sampling function.---}
In practice, the convergence of any sampling-based estimator is determined by its variance. The number of samples $M_{\rm s}$ required to achieve a desired relative error $\varepsilon$ scales as $M_{\rm s}\propto \sigma^2/(\varepsilon^{2} Z^2_1)$, where $\sigma^2$ is the variance of the estimator. Therefore, minimizing this variance is equivalent to minimizing the total computational cost. The JE estimator suffers from notoriously slow convergence precisely because its variance is large, especially at low temperatures \cite{Jarzynski_2006_PRE, Xiao_2014_PRE}. To address this, we propose replacing the conventional Gibbs initial distribution with a tailored nonequilibrium one. As shown below, this approach dramatically suppresses estimator variance, especially in the low-temperature regime.

From Eq.~\eqref{eq:Z_1}, the IPFs of $H_1$ can be estimated as
\be
Z_1(\beta) = \braket{z_{m,n}(\beta)} \approx Z_{\rm est}(\beta) \equiv \frac{1}{M_{\rm s}} \sum z_{m,n}(\beta)\, ,
\label{eq:Estimator}
\ee
where $\braket{\cdot}$ denotes averaging over trajectories $\ket{\psi^0_m}\to \ket{\psi^1_n}$ sampled with the probability $P_{n|m} P_m$, and $z_{m,n}(\beta) = \exp(-\beta E^1_n)/P_m$ is the corresponding random variable.
While the equality holds exactly as $M_{\rm s}\to\infty$, finite sampling necessitates clever design of the initial distribution $P_m$ and transition probability $P_{n|m}$ to minimize estimator variance.
Our protocol thus operates as a hybrid classical-quantum algorithm, i.e., the initial state is randomly chosen from $P_m$, a classical distribution designed to minimize statistical fluctuations in $Z_{\rm est}$; the transition $\ket{\psi^0_m}\to \ket{\psi^1_n}$ is governed by $P_{n|m}$, which is generated by the quantum dynamics of RQA.
For the latter, it is known that increasing the evolution time $\tau$ can reduce the required $M_{\rm s}$ \cite{Oberhofer_2005}.
However, practical limitations---notably finite coherence time in quantum hardware---constrain $\tau$, demanding a balance between quantum resource allocation and sampling efficiency. This raises the central question: what choice of $P_m$ minimizes the variance of $Z_{\rm est}$ for a given RQA protocol with finite $\tau$?

\begin{figure*}[htp!]
\centering
\includegraphics[width=1\textwidth]{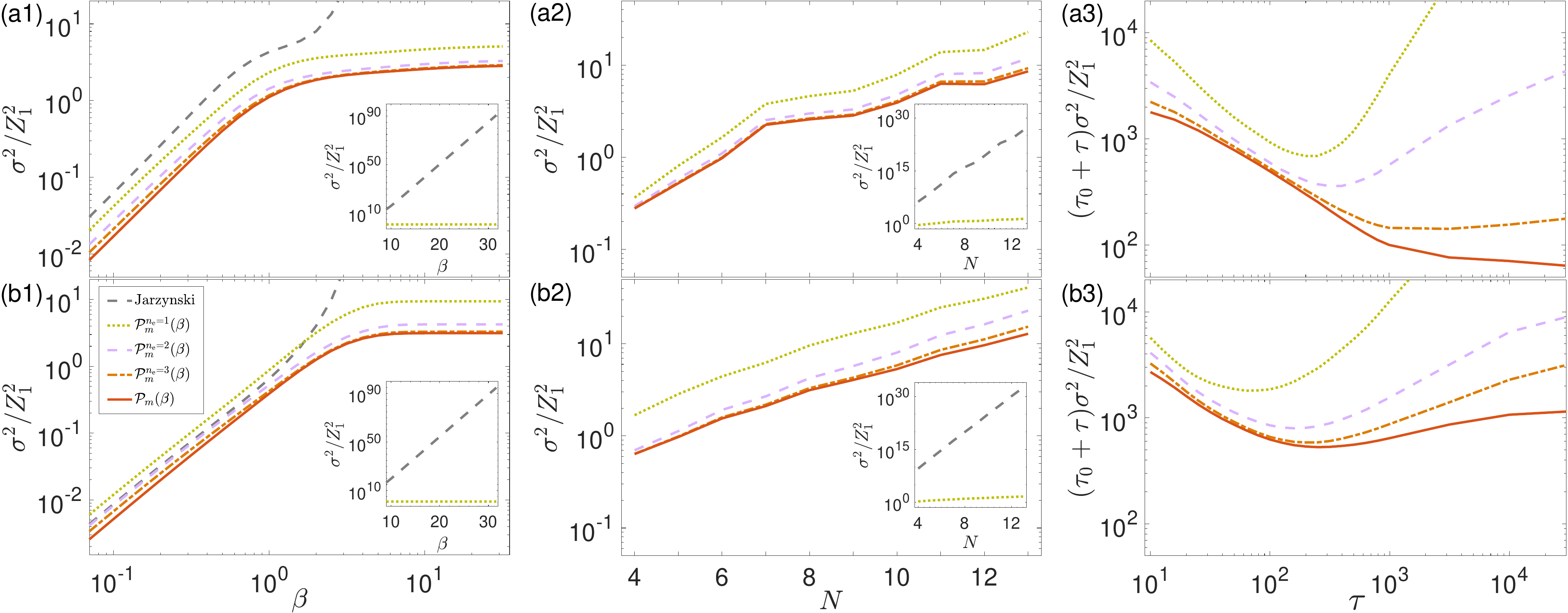}
\caption{
Performance evaluation of our protocol for the SK spin-glass (a1-a3) and 3-SAT (b1-b3) models.
The red solid line indicates the optimal sampling function $\mathcal{P}_m(\beta)$.
And the approximations $\mathcal{P}^{n_{\rm e}}_m(\beta)$ for $n_{\rm e}=1,2,3$ are depicted as follows: $n_{\rm e}=1$ (light green dotted line), $n_{\rm e}=2$ (purple dashed line), and $n_{\rm e}=3$ (orange dash-dotted line).
Results from the JE-based approach (gray dashed line) are included for comparison.
Here, we have chosen the system size $N=8$ for (a1, b1, a3, b3), the inverse temperature $\beta=10$ for (a2, b2, a3, b3), and the initialization and measurement time $\tau_0=100$ for (a3, b3).
All data are averaged over $10^3$ Hamiltonian instances.
}
\label{fig:Fig_3}
\end{figure*}

To address this problem, we examine the variance of the estimator $z_{m,n}(\beta)$, defined as \cite{OurSM}
\bea
\sigma^2(\beta)
&=& \sum\nolimits^D_{m,n=1} \left[z_{m,n}(\beta) - Z_1(\beta)\right]^2 P_{n|m} P_m \nn \\
&=& \sum\nolimits^D_{m=1}\frac{\mu_{m}(\beta)}{P_m}-Z^2_1(\beta)\, .
\label{eq:sigma}
\eea
Here, $\mu_{m}(\beta) = \sum^D_{n=1}\exp(-2\beta E^1_n)P_{n|m}$.
Guided by importance sampling principles \cite{Landau_2021}, we seek to minimize $\sigma^2(\beta)$ by optimizing $P_m$.
A direct constrained minimization yields the theoretically optimal distribution $P_m = \mathcal{P}_m(\beta)\equiv \sqrt{ \mu_{m}(\beta)}/\mathcal{N}_\mathcal{P}(\beta)$, where $\mathcal{N}_\mathcal{P}(\beta)$ ensures normalization \cite{OurSM}.
While $\mathcal{P}_m(\beta)$ guarantees the minimal variance, its direct computation is infeasible for finite $\tau$ in practical implementations. To resolve this, we develop an efficient approximation scheme for $\mathcal{P}_m(\beta)$ as follows.

The simplest approximation replaces $\mathcal{P}_m(\beta)$ with a Gibbs distribution $P^{\rm G}_m(\alpha)$, parameterized by a variational inverse temperature $\alpha$. For a fixed $\beta$, we optimize $\alpha$ to minimize $\sigma^2(\beta)$. This reduces to solving the minimization problem $\min_\alpha \sum^D_{m=1}\mu_m(\beta)/P^{\rm G}_m(\alpha)$, and from which we identify the unique optimal $\alpha$ that satisfies \cite{OurSM}
\be
 \frac{N}{1+\exp(\alpha)} = \sum\nolimits^N_{E=0} E\cdot Q_E(\alpha;\beta)\, .
 \label{eq:Alpha}
\ee
Here, $Q_E(\alpha;\beta) \propto \exp(\alpha E) \sum_{m\in I_E} \mu_m(\beta)$ is a distribution, and $I_E=\{m|E^0_m=E\}$ groups degenerate initial states $\ket{\psi^0_m}$ sharing the same energy $E$ [Fig.~\ref{fig:Fig_2} (a)].
Numerical results for the SK spin glass [Fig.~\ref{fig:Fig_2} (b)] reveal two key features: (i) As $\beta$ increases, $\alpha$ rises initially before saturating to a $\tau$-dependent value; (ii) Longer $\tau$ increases $\alpha$, concentrating $P^{\rm G}_m(\alpha)$ and reducing the computational effort to estimate $Z_1(\beta)$, which is consistent with previous works \cite{Oberhofer_2005}.
These behaviors starkly contrast with JE, which fixes $\alpha\equiv\beta$.
JE's rigidity leads to inefficient sampling in low temperatures, as high-energy states are inadequately probed.
Similar trends have been observed in other complex systems, such as the 3-SAT \cite{OurSM}, underscoring the generality of this approach.

While $P^{\rm G}_m(\alpha)$ outperforms $P^{\rm G}_m(\beta)$, determining $\alpha$ via Eq.~\eqref{eq:Alpha} for large system sizes remains challenging.
This issue can be mitigated based on three key observations.
Firstly, $P^{\rm G}_m(\alpha)$, while simpler, reasonably approximates the optimal distribution $\mathcal{P}_m(\beta)\propto \sqrt{ \mu_{m}(\beta)}$ \cite{OurSM}. Its defining property, $P^{\rm G}_{m\in I_{E+1}}(\alpha) = \exp(-\alpha)P^{\rm G}_{m\in I_{E}}(\alpha)$, leads to the approximation
\be
\bar{\nu}_{E+1}(\beta) \approx \exp(-\alpha)\bar{\nu}_{E}(\beta)\, ,
\label{eq:Mean_mu}
\ee
where $\bar{\nu}_{E}(\beta) = \sum_{m\in I_{E}}\sqrt{ \mu_{m}(\beta)}/\binom{N}{E}$ averages $\sqrt{ \mu_{m}(\beta)}$ over the $E^0_m=E$ degenerate subspace.
Secondly, combining Eq.~\eqref{eq:Mean_mu} and the definition of $Q_E$, we find that $Q_E$ decays exponentially with large $E$ [see Fig.~\ref{fig:Fig_2} (b), inset]. Consequently, only contributions from low-energy subspaces ($E\ll N$) significantly affect the expectation value on the RHS of Eq.~\eqref{eq:Alpha}.
Thirdly, for low-energy initial states ($m\in I_E$, $E\ll N$), we compute $\mu_{m}(\beta) = \sum^D_{n=1}\exp(-2\beta E^1_n)P_{n|m}$ using experimentally sampled trajectories $\ket{\psi^0_m}\to \ket{\psi^1_n}$ \cite{OurSM}. Crucially, these trajectories are reused to calculate $Z_1(\beta)$ after determining the sampling function $P_m$, incurring negligible additional computational costs.

Guided by these findings, we propose a family of sampling functions $\mathcal{P}^{n_{\rm e}}_m(\beta)\,(n_{\rm e} = 1,2,\dots,N)$ to approximate $\mathcal{P}_m(\beta)$. The construction involves four main steps as detailed in the Supplemental Material \cite{OurSM}.
Firstly, we introduce the auxiliary variables $\{\nu_{m}(\beta)\}|_{m=1,2,\dots,D}$ and set $\nu_{m}(\beta) = \sqrt{\mu_{m}(\beta)}$ for $m\in I_{E\leq n_{\rm e}}$. These $\mu_{m}(\beta)$ can be obtained experimentally via trajectory sampling as mentioned earlier.
Secondly, for $m\in I_{E > n_{\rm e}}$, we iteratively construct $\nu_{m}(\beta)$ via the ansatz
\be
\nu_{m\in I_{E+1}}(\beta) = \frac{\exp(-\alpha)}{E+1} \sum\nolimits_{m' \in J_{E}(m)}\nu_{m'}(\beta)\, ,
\label{eq:Ansatz}
\ee
where $\alpha$ is a parameter determined self-consistently, $J_{E}(m)= \{m'|m'\in I_E \,, m\in I_{E+1} \,, D_{\rm H}(m',m)=1\}$ contains states $m'\in I_E$ adjacent to $m$. Here, $D_{\rm H}(m',m)$ is the Hamming distance (i.e., the number of different spins) between $\ket{\psi^0_{m'}}$ and $\ket{\psi^0_m}$ [Fig.~\ref{fig:Fig_2} (a)].
This ensures consistency with the exponential decay relation in Eq.~\eqref{eq:Mean_mu}.
Thirdly, we approximate $\mu_{m}(\beta) \approx \nu^2_{m}(\beta)$ for $m\in I_{E>n_{\rm e}}$. Substituting all $\mu_{m}(\beta)$ into Eq.~\eqref{eq:Alpha} yields a closed-form solution for $\alpha$.
Finally, the $n_{\rm e}$-th order approximation is given by $\mathcal{P}^{n_{\rm e}}_m(\beta)\propto\nu_{m}(\beta)$, normalized over all $m$.
As shown in Fig.~\ref{fig:Fig_2} (c), we find that increasing $n_{\rm e}$ significantly improves fidelity of $\mathcal{P}_m(\beta)$, particularly at low temperatures. Besides, the approximation remains robust as system size $N$ grows [Fig.~\ref{fig:Fig_2} (c), inset], with significant error reduction even for modest $n_{\rm e}$.


{\it  Results.---}
To assess the efficacy of our protocol, we employ the relative estimator variance $\sigma^2/Z^2_1$ as the primary performance metric, since it directly determines the computational resources required for a given precision.
Numerical results for the SK spin glass and 3-SAT models, presented in Fig.~\ref{fig:Fig_3}, demonstrate a profound advantage over the standard JE-based approach.
Notably, in the low-temperature regime, the variance of our protocol---even when using the simplified sampling function $\mathcal{P}^{n_{\rm e}=1}_m(\beta)$---saturates, whereas the JE-based method exhibits exponential divergence [Fig.~\ref{fig:Fig_3} (a1, b1), insets].
Given that estimating IPFs is $\#$P-hard, quantum computers cannot reasonably be expected to solve this task in polynomial time.
Performance is therefore quantified by the exponential scaling of the variance with system size, $\sigma^2/Z^2_1\sim D^\gamma=2^{\gamma N}$. At a representative low temperature ($\beta=10$), we analyze this scaling for our protocol versus the JE baseline [Fig.~\ref{fig:Fig_3} (a2, b2)]. For the SK spin glass, our protocol with $\mathcal{P}^{n_{\rm e}=1}_m(\beta)$ achieves a scaling exponent of $\gamma \approx 0.446$, whereas the JE approach yields $\gamma \approx 6.993$ for the hard 3-SAT instances, the exponents are $\gamma \approx 0.502$ for our protocol versus $\gamma \approx 8.451$ for JE. This reduction in the scaling exponent by over an order of magnitude corresponds to a substantial reduction in the exponential scaling of the total computational cost (i.e., the number of experimental runs) required to solve a problem of a given size $N$.
A full complexity analysis of the scaling properties of $\mathcal{P}_m(\beta)$ further corroborates these results \cite{OurSM}.

Our protocol is not only powerful but also practical for near-term implementation. Because the number of samples needed scales with the variance, we define the total resource cost as $\mathcal{M}_{\rm s} = (\tau_0+\tau)\sigma^2$, where $\tau_0$ denotes the initialization and measurement time per sample.
While a longer evolution time $\tau$ generally reduces the variance $\sigma^2$, there is a trade-off. As shown empirically in Fig.~\ref{fig:Fig_3} (a3, b3), this trade-off leads to an optimal, finite $\tau$ that minimizes the total cost. This confirms that the most efficient implementation of our protocol operates in a non-adiabatic regime, perfectly aligning with the limited coherence times of NISQ devices. This performance represents a substantial advantage over both the JE baseline and other known approaches in the challenging low-temperature regime. While JE can fail due to rare-event-dominated statistics, classical MCMC methods can become severely hindered by extremely long (in unfavorable cases, effectively exponential) equilibration times in rugged spin-glass landscapes \cite{BenArous2018,Gheissari2019}; enhanced variants such as annealed importance sampling or generalized-ensemble schemes may improve exploration but can still struggle with low-temperature barriers \cite{Dayal2004WLspinGlass}. More advanced quantum algorithms based on phase estimation \cite{Poulin_2009_PRL} or DQC1 \cite{Chowdhury_2021_PRA,Jackson_2023_PRA} require fault-tolerant hardware and are thus impractical for near-term implementation. A detailed complexity comparison is provided in the Supplemental Material \cite{OurSM}.

{\it Conclusion and discussion.---}
We have introduced an efficient, hardware-native protocol for estimating IPFs, a foundational $\#$P-hard problem. By synergizing RQA with optimized nonequilibrium initial distributions, our method circumvents the fundamental limitations of conventional approaches based on JE. The protocol's key achievements are the saturation of estimator variance at low temperatures and a substantial improvement in computational efficiency and scalability. These results highlight the power of harnessing non-equilibrium quantum dynamics to tackle classically difficult problems.
The underlying algorithmic paradigm---the synergistic use of optimized non-equilibrium initial states and non-adiabatic quantum dynamics---is a generalizable strategy with broad relevance beyond spin glasses.
In realistic implementations, environmental noise can substantially affect quantum annealing dynamics \cite{Passarelli_2020_PRA, Passarelli_2022_PRA}, and a comprehensive noise-aware validation---given ongoing debates over appropriate modeling \cite{Le_2025ar}---is deferred to future work.
As quantum hardware matures, such strategies could unlock new insights into the complex energy landscapes of problems in materials science, protein folding, and machine learning \cite{Doga_2024, Cerezo_2022_NCS}. Our work provides a concrete blueprint for leveraging the unique capabilities of near-term quantum processors \cite{Albash_2018_PRX, Hauke_2020_IOP, Miessen_2024_PRXQuantum, Guo_2024_Nature, Kihwan_2025_PRL, Qiu_2020_PRXQ, Ebadi_2022_Science}, bridging the gap between theoretical quantum enhancement and impactful real-world applications.

\begin{acknowledgments}
{\it Acknowledgement.---}
Zhiyuan Yao acknowledges support by National Natural Science Foundation of China (12304288, 12247101).
Xingze Qiu acknowledges support from the Fundamental Research Funds for the Central Universities and Shanghai Science and Technology project (24LZ1401600).
\end{acknowledgments}

{\it Data availability.---} The source code and data supporting the findings of this article are openly available at \cite{LiYaoQiu2026}.

\bibliography{References}

\end{document}


\title{Supplemental Material for ``Quantum Annealing Algorithms for Estimating Ising Partition Functions''}

\author{Haowei Li}
\affiliation{School of Physics Science and Engineering, Tongji University, Shanghai 200092, China}
\affiliation{Laboratory of Quantum Information, University of Science and Technology of China, Hefei 230026, China}
\affiliation{Institute for Advanced Study, Tsinghua University, Beijing 100084, China}

\author{Zhiyuan Yao}
\email{yaozy@lzu.edu.cn}
\affiliation{Lanzhou Center for Theoretical Physics, Key Laboratory of Theoretical Physics of Gansu Province, Key Laboratory of Quantum Theory and Applications of MoE, Gansu Provincial Research Center for Basic Disciplines of Quantum Physics, Lanzhou University, Lanzhou 730000, China}

\author{Xingze Qiu}
\email{xingze@tongji.edu.cn}
\affiliation{School of Physics Science and Engineering, Tongji University, Shanghai 200092, China}

\maketitle

\renewcommand{\theequation}{S\arabic{equation}}
\renewcommand{\thefigure}{S\arabic{figure}}

In this Supplemental Material, we provide detailed discussions on the following topics:  the construction of the 3-SAT Hamiltonian, the variance analysis of the sampling process, the Gibbs distribution approximation for the sampling function, the protocol for determining the sampling function, the complexity analysis of the proposed protocol, and comparison of our protocol against several other algorithms.

\section{I. The 3-SAT Hamiltonian}
In this section, we introduce the 3-SAT problem and its mapping to the Ising Hamiltonian.

The 3-satisfiability (3-SAT) problem is a canonical NP-complete problem of significant importance in both classical and quantum computations \cite{Monasson1999, Mezard2002, 3SAT2002}. A random 3-SAT formula \( F \) consists of \( M \) logical clauses \( \{C_m\}_{m=1, \ldots, M} \) defined over a set of \( N \) Boolean variables \( \{x_i \in \{0, 1\}\}_{i=1, \ldots, N} \), where \( 0 \) typically represents FALSE and \( 1 \) represents TRUE. Each clause \( C_m \) involves three distinct Boolean variables, chosen randomly and uniformly from the \(N\) variables. These variables are joined by logical OR operations (\( \vee \)), and a typical form of a clause reads \( C_m = (x_i \vee \overline{x_j} \vee x_k) \) where $\overline{x}$ denotes the negation of $x$. The overall formula \( F \) is the conjunction (logical AND) of all $M$ clauses, \( F = \wedge_{m=1}^M C_m \), which evaluates to TRUE if and only if all clauses are simultaneously satisfied. The task of the 3-SAT problem is to find solutions $\{x_{i}\}$ that makes a particular conjunction normal form $F$ true, i.e., each clause evaluated to be true.

To construct a 3-SAT problem, we focus on instances constructed around a planted solution. We assume the existence of at least one satisfying assignment \( \{x_i^0\} \), where \( x_i^0 \in \{0, 1\} \) for \( i = 1, \ldots, N \). This assignment serves as the planted solution. For each of the $M$ clauses, three distinct indices \( i, j, k \in \{1, \ldots, N\} \) are drawn randomly and independently. The clause structure is then chosen based on the values of the variables in the planted solution \( \{x_i^0\} \), ensuring that \( \{x_i^0\} \) satisfies the clause. Specifically, one of the following seven types of clauses (relative to the planted solution being satisfied) is selected according to the specified probabilities:

\begin{itemize}
	\item Type ``0": Clauses have the form \( (l_i \vee l_j \vee l_k) \) where none of the literals $l_i, l_j, l_k$ evaluate to FALSE under the planted solution with probability \( p_0 \).
	\item Type ``1": Clauses have one literal that evaluates to FALSE in the planted solution [e.g. \( (\overline{l_i} \vee l_j \vee l_k) \) if $l_i$ is TRUE]. There are three such possibilities, each chosen with probability \( p_1 \).
	\item Type ``2": Clauses have two literals that evaluate to FALSE in the planted solution [e.g. \( (\overline{l_i} \vee \overline{l_j} \vee l_k) \) if $l_i, l_j$ are TRUE]. Three such possibilities exist and each is chosen with probability \( p_2 \).
\end{itemize}
Here, $l_i$ represents either $x_i$ or $\overline{x_i}$. The probabilities are normalized so that \( p_0 + 3p_1 + 3p_2 = 1 \). This constructive method guarantees that the planted assignment \( \{x_i^0\} \) satisfies all clauses. In our study, we use  parameters \( \alpha \equiv M/N \approx 4.25 \), \( p_0 = 1/7 \), \( p_1 = 1/14 \), and \( p_2 = 3/14 \). These parameters are known to generate particularly hard 3-SAT instances \cite{3SAT2002}.

This above 3-SAT problem can be mapped to a classical Ising Hamiltonian of \( N \) spins \( \sigma_z^i \in \{+1, -1\} \) by identifying TRUE ($x_i=1$) to spin-up ($\sigma_z^i=+1$) and FALSE ($x_i=0$) to spin-down ($\sigma_z^i=-1$). An Ising Hamiltonian for the 3-SAT problem can be constructed in the following way \cite{3SAT2002}
\begin{equation}
    H_1 =  \frac{1}{8} \sum_{m=1}^{M}\left(1-c_{mi} \sigma_z^i\right)\left(1-c_{mj} \sigma_z^j\right)\left(1-c_{mk} \sigma_z^k\right)
    = \frac{\alpha}{8} N - \sum_{i=1}^N F_i \sigma_z^i - \sum_{i<j} J_{ij} \sigma_z^i \sigma_z^j - \sum_{i<j<k} K_{ijk} \sigma_z^i \sigma_z^j \sigma_z^k,
\end{equation}
where the parameter $c_{mi}$ takes $1~(-1)$ if $l_{i} = x_{i}~(\overline{x_{i}})$ and zero if $x_{i}$ and its negation are absent in $C_{m}$, and the random couplings are given by $F_i = \frac{1}{8} \sum_m c_{mi}$, $J_{ij} = -\frac{1}{8} \sum_m c_{mi} c_{mj}$, and $K_{ijk} = \frac{1}{8} \sum_m c_{mi} c_{mj} c_{mk}$, respectively. By construction, the eigenenergies of the Hamiltonian are nonnegative. The coefficient $1/8$ is for normalization purposes so that the energy of the Hamiltonian equals the number of unsatisfied clauses, and solutions to the 3-SAT problem are in one-to-one correspondence with the zero-energy ground states.

\section{II. Variance analysis}

In this section, we provide a detailed analysis of the variance of the partition function estimator $z_{m,n}(\beta)$ and derive the form of the optimal sampling function $\mathcal{P}_m(\beta)$ that minimizes this variance.

The variance of $z_{m,n}(\beta)$, defined in Eq.~(4) of the main text, is derived as follows:
\begin{equation}
	\begin{aligned}
		\sigma^2(\beta)&=\langle(z_{m,n}(\beta)-Z_1(\beta))^2\rangle=\sum_{m,n=1}^DP_mP_{n|m}\left(\frac{e^{-\beta E^1_n}}{P_m}-Z_1(\beta)\right)^2\\&=\sum_{m,n=1}^DP_mP_{n|m}\left( \frac{e^{-2\beta E^1_n}}{P_m^2}-2Z_1(\beta)\frac{e^{-\beta E^1_n}}{P_m}+Z_1^2(\beta)\right)\\&=\sum_{m=1}^D \frac{\sum_{n=1}^D e^{-2\beta E_n^1} P_{n|m}}{P_m}-2Z_1(\beta)\sum_{n=1}^De^{-\beta E_n^1}\sum_{m=1}^D P_{n|m}+Z_1^2(\beta)\\&=\sum_{m=1}^D \frac{\mu_m(\beta)}{P_m}-Z_1^2(\beta),
	\end{aligned}
	\label{sigma2z}
\end{equation}
where we have defined $\mu_m(\beta)=\sum_{n=1}^D e^{-2\beta E_n^1} P_{n|m}$ in the derivation to match the form in Eq.~(4) of the main text.

To find the optimal sampling function $P_m=\mathcal{P}_m(\beta)$ that minimizes the variance in Eq.~(\ref{sigma2z}) under the normalization constraint $\sum_{m=1}^D P_m = 1$, we use the method of Lagrange multipliers and minimize the Lagrangian function $\mathcal{L}$,
\begin{equation}
	\mathcal{L}(P_1, P_2, \dots, P_D, \lambda) = \sum_{m=1}^D \frac{\mu_m(\beta)}{P_m} - Z_1^2(\beta) + \lambda \left( \sum_{m=1}^D P_m - 1 \right) \, .
\end{equation}
The conditions for a minimum are:
\begin{align}
	\frac{\partial \mathcal{L}}{\partial P_m} &= -\frac{\mu_m(\beta)}{P_m^2} + \lambda = 0, \quad \text{for } m = 1,2, \dots, D, \\
	\frac{\partial \mathcal{L}}{\partial \lambda} &= \sum_{m=1}^D P_m - 1 = 0.
\end{align}
From the first equation, we have $P_m = \sqrt{\mu_m(\beta) / \lambda}$. The constant $\lambda$ is determined by the normalization constraint so that $\sqrt{\lambda} = \sum_{m=1}^D \sqrt{\mu_m(\beta)}$. Therefore, the optimal sampling function $\mathcal{P}_m(\beta)$ that minimizes the variance is:
\begin{equation}
	P_m = \mathcal{P}_m(\beta) = \frac{\sqrt{\mu_m(\beta)}}{\sum_{m^\prime=1}^D \sqrt{\mu_{m^\prime}(\beta)}}.
	\label{eq:P_optimal}
\end{equation}
Substituting this optimal $P_m=\mathcal{P}_m(\beta)$ back into Eq.~(\ref{sigma2z}) yields the minimal variance:
\begin{equation}
	\sigma_{\mathrm{min}}^2(\beta) = \left( \sum_{m=1}^{D} \sqrt{\mu_m(\beta)} \right)^2 - Z_1^2(\beta).
	\label{varmin}
\end{equation}

\section{III. Gibbs distribution approximation of the sampling function}
\begin{figure}[tbp]
	\includegraphics[width=0.65\textwidth]{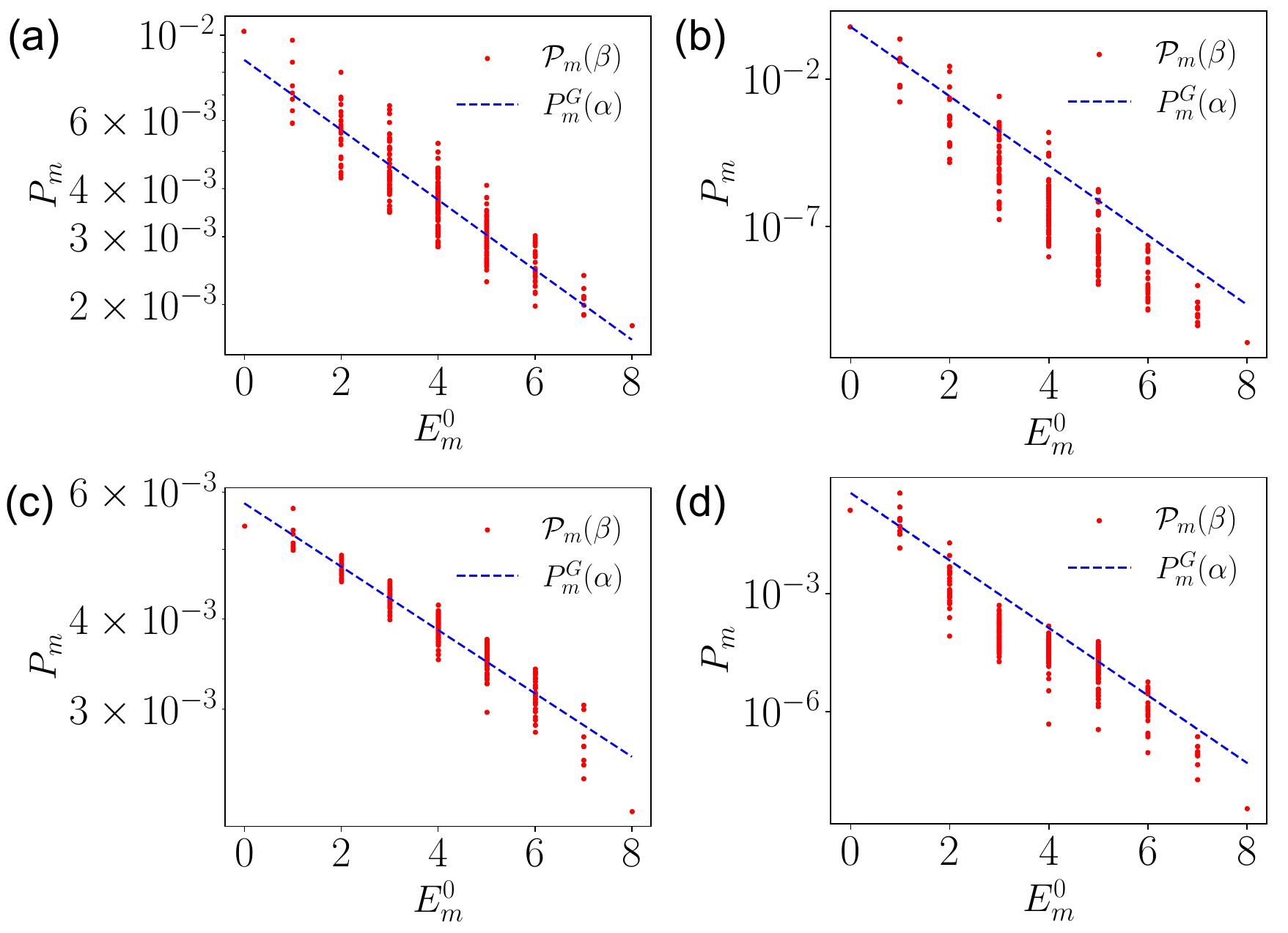}
	\caption{Comparison of the optimal sampling function $\mathcal{P}_m(\beta)$ (red solid dots) with the Gibbs approximation $P_m^G(\alpha)$ (blue dashed line). The parameter $\alpha$ for $P_m^G(\alpha)$ is determined by minimizing the KL divergence from $\mathcal{P}_m(\beta)$ to $P_m^G(\alpha)$. Panels (a) and (b) are for the SK model, while (c) and (d) are for the 3-SAT model. Panels (a) and (c) use $\beta=0.1$, while (b) and (d) use $\beta=10$. For all plots, $N=8$, $\tau=100$, and results are shown for a typical Hamiltonian instance.}
	\label{figS1}
\end{figure}

The optimal function $\mathcal{P}_m(\beta)$ Eq.~(\ref{eq:P_optimal}) minimizes the variance $\sigma^2(\beta)$, but requires the knowledge of all $\mu_m(\beta)$, which is in general unavailable. In this section, we approximate it using a Gibbs distribution. This provides insights for designing a feasible protocol to determine the sampling function in practice.

Fig.~\ref{figS1} shows the optimal sampling function $\mathcal{P}_m(\beta)$ as a function of the eigenenergy $E_m^0$ of $H_{0}$. We observe that $\mathcal{P}_m(\beta)$ generally decreases exponentially with $E_m^0$. This observation motivates approximating the optimal sampling function $\mathcal{P}_m(\beta)$ with a Gibbs distribution based on the initial Hamiltonian $H_0$:
\begin{equation}
	P_m^G(\alpha) = \frac{e^{-\alpha E_m^0}}{Z_0(\alpha)},
	\label{eq:P_Gibbs}
\end{equation}
where the normalization factor $Z_0(\alpha) = \sum_{m=1}^{D} e^{-\alpha E_m^0}$ is the partition function of $H_0 = \sum_{i=1}^N (1 - \sigma_z^i)/2$ at inverse temperature $\alpha$. Since the Hamiltonian is a sum of independent one-body ones, the partition function can be easily written down
\begin{equation}
	Z_0(\alpha)=\sum_{m=1}^{D}e^{-\alpha E_m^0}=(1+e^{-\alpha})^N,
\end{equation}

We now consider the variance of $z_{m,n}(\beta)$ when using the Gibbs sampling function $P_m^G(\alpha)$, denoted as $\sigma^{2}(\alpha; \beta)$. Substituting $P_m^G(\alpha)$ into the last expression of Eq.~(\ref{sigma2z}), we get:
\begin{equation}
	\sigma^{2}(\alpha; \beta) = \sum_{m=1}^{D} \frac{\mu_m(\beta)}{P_m^G(\alpha)} - Z_1^2(\beta) = Z_0(\alpha) \sum_{m=1}^D \mu_m(\beta) e^{\alpha E^0_m} - Z_1^2(\beta).
	\label{mu2balp}
\end{equation}
The optimal $\alpha$ that minimizes $\sigma^{2}(\alpha; \beta)$ has then to obey
\begin{equation}
	\frac{\partial \sigma^{2}(\alpha; \beta)}{\partial \alpha} = \frac{Z_0(\alpha)}{1+e^\alpha}\sum_m \left[E^0_m(1+e^\alpha) -N \right] \mu_m(\beta)e^{\alpha E^0_m} = 0.
	\label{alpeq1}
\end{equation}
This condition can be rearranged into the following self-consistent equation for $\alpha$,  Eq.~(5) in the main text,
\begin{equation}
	\frac{N}{1 + e^\alpha} = \frac{\sum_m E^0_m \mu_m(\beta) e^{\alpha E^0_m}}{\sum_m \mu_m(\beta) e^{\alpha E^0_m}} \equiv \sum_{E=0}^N E \cdot Q_E(\alpha;\beta),
	\label{alpeq2}
\end{equation}
Here $Q_E(\alpha;\beta)$ is a probability distribution over energy $E$ that can be written as
\begin{equation}
	Q_E(\alpha;\beta) = \frac{\sum_{m \in I_E} \mu_m(\beta) e^{\alpha E}}{\sum_{E'=0}^N \sum_{m' \in I_{E'}} \mu_{m'}(\beta) e^{\alpha E'}}, \quad \text{where } I_E = \{m \mid E^0_m = E\}.
\end{equation}

Furthermore, we examine the second derivative of $\sigma^2(\alpha; \beta)$ to ensure that the solution corresponds to a minimum:
\begin{align}
	\frac{\partial^2 \sigma^2(\alpha; \beta)}{\partial \alpha^2} &= \sum_m \mu_m(\beta) \frac{\partial^2}{\partial \alpha^2} \left( Z_0(\alpha) e^{\alpha E^0_m} \right) \nonumber \\
	&= \sum_m \mu_m(\beta) \frac{\partial}{\partial \alpha} \left[ \frac{Z_0(\alpha)}{1+e^\alpha} e^{\alpha E^0_m} \left( E^0_m (1+e^\alpha) - N \right) \right] \nonumber \\
	&= \sum_m \mu_m(\beta) Z_0(\alpha) e^{\alpha E^0_m} \left\{ \left( E^0_m - \frac{N}{1+e^\alpha} \right)^2 + \frac{N e^\alpha}{(1+e^\alpha)^2} \right\} \nonumber \\
	&= \frac{Z_0(\alpha)}{(1 + e^\alpha)^2} \sum_m \mu_m(\beta) e^{\alpha E^0_m} \left\{ [N - (1 + e^\alpha) E^0_m]^2 + N e^\alpha \right\}. \label{eq:second_deriv_sigma2}
\end{align}
Since all terms in the sum are nonnegative, the second derivative is positive. This confirms that the $\alpha$ satisfying Eq.~(\ref{alpeq1}) corresponds to a global minimum of the variance $\sigma^2(\alpha; \beta)$ when using the Gibbs approximation $P_m^G(\alpha)$.

\begin{figure}[tbp]
	\includegraphics[width=0.65\textwidth]{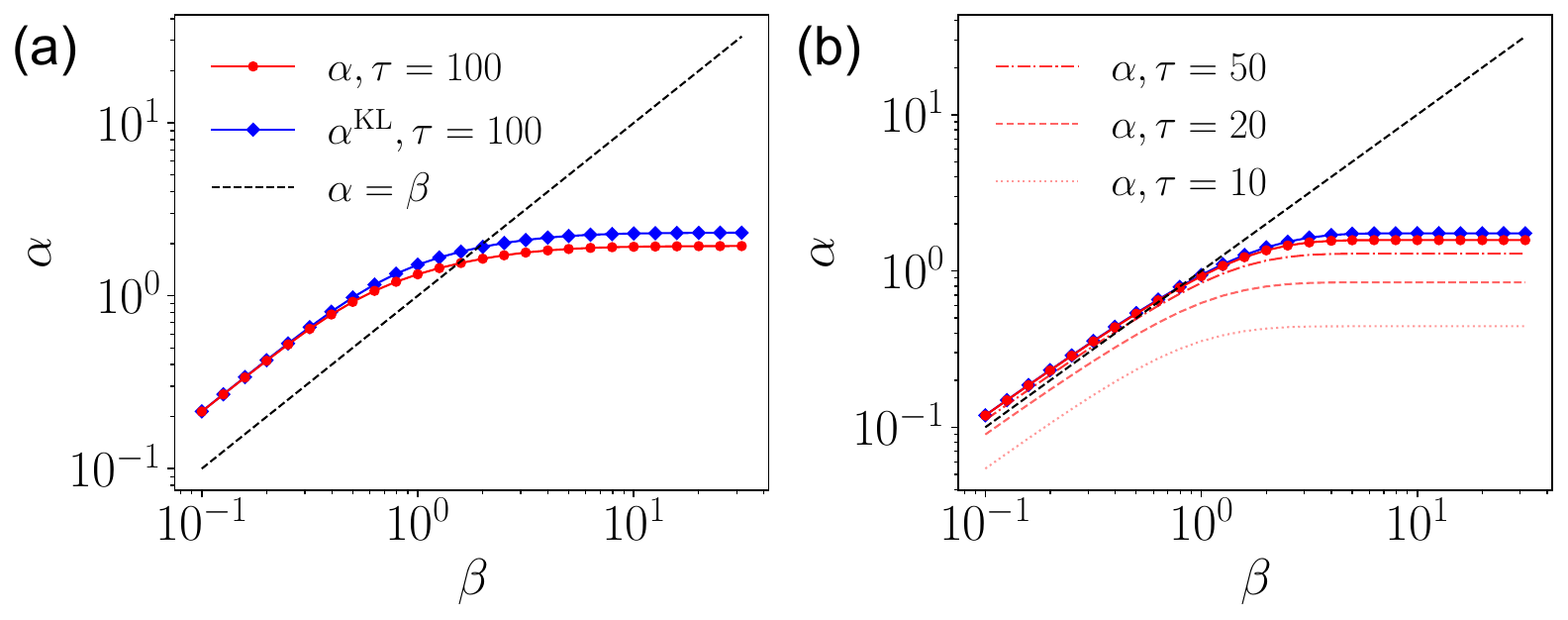}
	\caption{Relationship between the optimal $\alpha$ of the Gibbs approximation $P_m^G(\alpha)$ and the inverse temperature $\beta$. Red circles: $\alpha$ determined by minimizing the variance $\sigma^2(\beta)$ using the self-consistent Eq.~(\ref{alpeq2}) for $\tau=100$. Blue diamonds: $\alpha^{\mathrm{KL}}$ determined by minimizing the KL divergence $D_{\mathrm{KL}}(\mathcal{P} || P^G)$ using Eq.~(\ref{alpKL}) for $\tau=100$. The black dashed line indicates $\beta = \alpha$. Panels (a) and (b) are for the SK and 3-SAT models, respectively. The light red dash-dotted, dashed, and dotted lines in (b) correspond to $\tau=50, 20, 10$ respectively for the 3-SAT model. Parameters: $N=8$, averaged over $1000$ Hamiltonian instances.}
	\label{figS2}
\end{figure}

Besides determining $\alpha$ from the minimal variance condition Eqs.~(\ref{alpeq1})--(\ref{alpeq2}), we can directly fit the optimal sampling function $\mathcal{P}_m(\beta)$ with the Gibbs distribution $P_m^G(\alpha)$. Our fitting is performed by minimizing the Kullback-Leibler (KL) divergence from the optimal distribution $\mathcal{P}_m(\beta)$ Eq.~(\ref{eq:P_optimal}) to the Gibbs approximation $P_m^G(\alpha)$ Eq.~(\ref{eq:P_Gibbs}):
\begin{equation}
	D_{\mathrm{KL}}(\mathcal{P} || P^G)=\sum_{m=1}^{D} \mathcal{P}_m(\beta)\ln\frac{\mathcal{P}_m(\beta)}{P_m^G(\alpha)}.
\end{equation}
Minimizing $D_{\mathrm{KL}}$ with respect to $\alpha$ gives the condition:
\begin{equation}
	\frac{\partial D_{\mathrm{KL}}(\mathcal{P} || P^G)}{\partial \alpha}
	= \frac{1}{1+e^\alpha}\sum_m \left[E^0_m(1+e^\alpha) -N \right] \mathcal{P}_m(\beta) = 0.
	\label{alpKL}
\end{equation}
We plot the fitted Gibbs distribution for several typical Hamiltonian instances as the blue lines in Fig.~\ref{figS1}.
This condition Eq.~(\ref{alpKL}) is similar but not identical to Eq.~(\ref{alpeq1}), as it involves $\mathcal{P}_m(\beta)$ instead of $\mu_m(\beta) e^{\alpha E_m^0} \propto \mathcal{P}_m^2(\beta)/P_m^G(\alpha)$. With our assumption that $\mathcal{P}_m(\beta)\approx P_m^G(\alpha)$, the values of $\alpha$ determined by these two methods are expected to be close. In Fig.~\ref{figS2}, we plot the optimal $\alpha$ as a function of $\beta$, determined using both the minimum variance condition Eq.~(\ref{alpeq1}) (red circles) and the minimum KL divergence condition Eq.~(\ref{alpKL}) (blue diamonds). The values obtained via these two methods exhibit the same trend and are quantitatively close in the high-temperature regime.

In the main text, we show the $\alpha-\beta$ relationship for the SK model under different evolution times $\tau$ in Fig.~2(b). There, we found that $\alpha$ first increases with increasing $\beta$ and then saturates in the low-temperature regime, and that the saturation value is larger for longer $\tau$. In Fig.~\ref{figS2}(b), we show the $\alpha-\beta$ dependence for the 3-SAT model under different $\tau$, and find a similar behavior.

\section{IV. Sampling function determination protocol}

\begin{figure}[tbp]
	\includegraphics[width=0.99\textwidth]{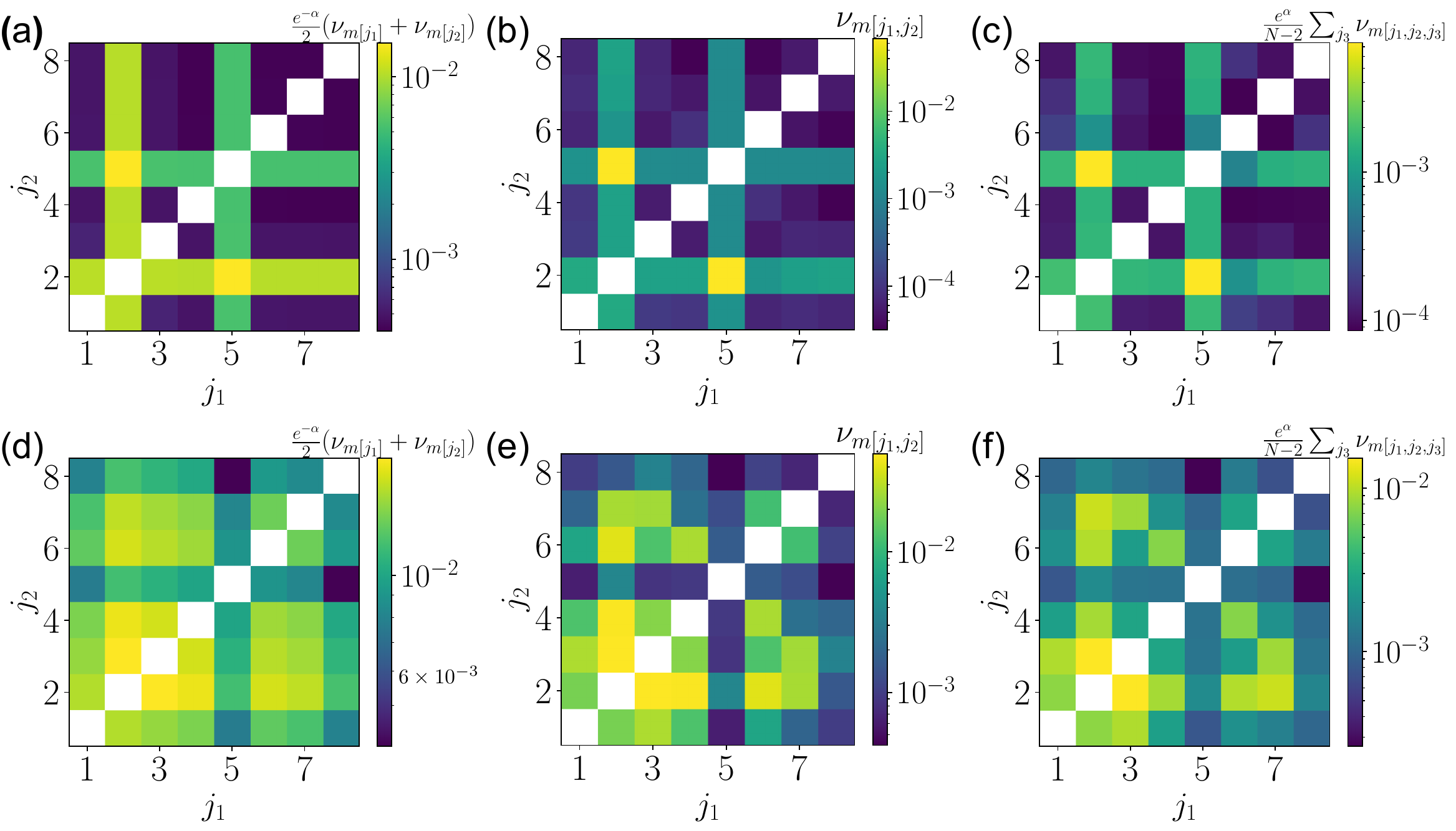}
	\caption{Numerical evidences supporting the locality assumption used in extrapolating $\hat{\nu}_m(\beta)$. Panels (a)--(c) are for the SK model, (d)--(f) for the 3-SAT model. The plots show distributions of quantities related to $\hat{\nu}_m(\beta) = \sqrt{\hat{\mu}_m(\beta)}$ for states $m$ with initial energies $E_m^0=1, 2, 3$. Here, $m[j_1, \dots, j_k]$ denotes the initial state obtained by flipping $k$ spins at sites $j_1, \dots, j_k$ relative to the ground state $|\psi_0^0\rangle$ (all spins up, $E^0=0$) of $H_0$. (a) and (d): Distributions of $\frac{e^{-\alpha}}{2}(\hat{\nu}_{m[j_1]} + \hat{\nu}_{m[j_2]})$, representing a prediction for $\hat{\nu}_{m[j_1, j_2]}$ (energy $E^0=2$) based on single-flip states ($E^0=1$). (b) and (e): Actual distributions of $\hat{\nu}_{m[j_1, j_2]}$ for double-flip states. (c) and (f): Distributions of $\frac{e^{\alpha}}{N-2}\sum_{j_3 \neq j_1, j_2} \hat{\nu}_{m[j_1, j_2, j_3]}$ that use information from related triple-flip states ($E^0=3$) to infer properties of the double-flip state. The statistical similarity between distributions in the same row supports the acceptability of the recursive relation Eq.~(\ref{nudrv}). Parameters: $N=8$, $\tau=100$, $\beta=10$. Results shown are for a typical Hamiltonian instance.
In all plots, the diagonal elements $j_1=j_2$ are excluded and not shown (white blocks).
	}
	\label{figS3}
\end{figure}


While the theoretical form of the optimal sampling function is known $\mathcal{P}_m(\beta) \propto \sqrt{\mu_m(\beta)}$, calculating $\mu_m(\beta)$ for all $m$ is infeasible.
Observing that $\mathcal{P}_m(\beta)$ decays exponentially with energy, much like a Gibbs distribution, we implement a sampling protocol that uses direct sampling for low-energy states and extrapolation for high-energy states.

\textbf{1. Presampling for Low-energy States:} First, a presampling stage is dedicated to  initial states $|\psi_m^0\rangle$ with low energies, $E_m^0 \le n_{\rm e}$. The goal of this stage is to gather data needed to construct the sampling function before the run of  partition function estimation. For each such state $m$, we execute the unitary time evolution $M_{\mathrm{ps}}(m)$ times. In the $i$-th run (starting from $|\psi_m^0\rangle$), we perform a measurement in the energy eigenbasis of $H_1$ and collect the measurement outcome $E_{n_i}^1$. We then estimate the quantity $\mu_m(\beta)$ using the sample average:
\begin{equation}
	\hat{\mu}_m(\beta) = \frac{1}{M_{\mathrm{ps}}(m)}\sum_{i=1}^{M_{\mathrm{ps}}(m)}\exp(-2\beta E_{n_i}^1).
	\label{eq:mu_hat}
\end{equation}
The estimate of $\nu_m(\beta) \equiv \sqrt{\mu_m(\beta)}$ is then $\hat{\nu}_m(\beta) = \sqrt{\hat{\mu}_m(\beta)}$. Note that the data collected, pairs of $(m, E_{n_i}^1)$, during this presampling stage can be reused later in the final partition function estimation step [Eq.~(3) in the main text]. The computational cost of this stage is limited because it only involves states with low initial energies, $E_m^0 \le n_{\rm e}$, and the number of such states $\sum_{E=0}^{n_{\rm e}} \binom{N}{E}$ is polynomial in $N$ for fixed $n_{\rm e}$.

\textbf{2. Extrapolation for High-energy States:} For initial state $|\psi_m^0\rangle$ with energy $E_m^0 = E+1 > n_{\rm e}$, we need to estimate $\nu_m(\beta)$ without direct simulation. Based on the observation (Sec. III) that the optimal distribution $\mathcal{P}_m(\beta) \propto \nu_m(\beta)$ decays approximately exponentially with $E_m^0$ (like $e^{-\alpha E_m^0}$), we propose a recursive relation to extrapolate $\hat{\nu}_m(\beta)$ for states with higher energies $E = n_{\rm e}+1, n_{\rm e}+2, \ldots, N$,
\begin{equation}
	\hat{\nu}_{m \in I_{E+1}}(\beta) \approx \frac{e^{-\alpha}}{E+1} \sum_{m^{\prime} \in J_E(m)} \hat{\nu}_{m^{\prime}}(\beta).
	\label{nudrv}
\end{equation}
Here, $I_E = \{ m \mid E_m^0 = E \}$ is the collection of degenerate levels of energy $E$, and $J_E(m) = \{ m^{\prime} \in I_E \mid D_{\mathrm{H}}(m^{\prime}, m)=1 \}$ is the set of states $m'$ in $I_E$ that differ from the state $m \in I_{E+1}$ by one local spin configuration (Hamming distance $D_{H}=1$), which we shall call neighbors of $m$ . The factor $e^{-\alpha}$ accounts for the expected exponential decay between energy levels $E$ and $E+1$. The term $\frac{1}{E+1}$ comes from the connectivity of the underlying hypercube graph (a state $m \in I_{E+1}$ has $E+1$ neighbors in $I_E$). The sum averages the $\hat{\nu}_{m'}$ values from these neighbors. The use of Hamming distance 1 neighbors reflects an assumption of locality. Fig.~\ref{figS3} provides numerical evidences supporting the validity of this recursive approach by showing that  $\hat{\nu}_m$ constructed from adjacent energy levels exhibit similar patterns, consistent with Eq.~(\ref{nudrv}).

\textbf{3. Calculating Averages of $\hat{\nu}_m(\beta)$ and $[\hat{\nu}_m(\beta)]^2$:} While Eq.~(\ref{nudrv}) could, in principle, determine all $\hat{\nu}_m(\beta)$, computing and storing them individually is infeasible due to the exponential size of the Hilbert space. However, the sampling probability $P_m \propto \hat{\nu}_m(\beta)$ becomes negligible for large $E_m^0$. Crucially, in determining the normalization factor of the sampling function and the parameter $\alpha$, we only need average quantities over energy shells $I_E$ for $E > n_{\rm e}$:
\begin{itemize}
	\item $\bar{\nu}_E(\beta) = \binom{N}{E}^{-1} \sum_{m\in I_E} \hat{\nu}_m(\beta)$, the average value of $\hat{\nu}_m(\beta)$ over states with energy $E$. Needed for the normalization factor $\sum_m \hat{\nu}_m(\beta) = \sum_E \binom{N}{E} \bar{\nu}_E(\beta)$ of the sampling function.
	\item $\bar{\mu}_E(\beta) = \binom{N}{E}^{-1} \sum_{m\in I_E} [\hat{\nu}_m(\beta)]^2$, the average value of $\hat{\mu}_m(\beta)$ over states with energy $E$. Needed to estimate the parameter $\alpha$ using the self-consistent equation Eq.~(\ref{alpeq2}) (see below).
\end{itemize}
Assuming the recursive relation Eq.~(\ref{nudrv}) holds on average and leveraging the symmetry among states within an energy shell $I_E$ (for the initial Hamiltonian $H_0$), we can derive relations for these averages for $E > n_{\rm e}$. Averaging Eq.~(\ref{nudrv}) over all $m \in I_{E+1}$ gives:
\begin{equation}
	\bar{\nu}_{E+1}(\beta) = e^{-\alpha} \bar{\nu}_{E}(\beta), \quad \text{for } E \ge n_{\rm e}.
	\label{nubar_rec}
\end{equation}
This implies $\bar{\nu}_E(\beta) = e^{-\alpha(E-n_{\rm e})} \bar{\nu}_{n_{\rm e}}(\beta)$ for $E > n_{\rm e}$.

To estimate $\bar{\mu}_E(\beta)$ for $E > n_{\rm e}$, we relate the average second moment within shell $I_E$ to the moments within the base shell $I_{n_{\rm e}}$. By repeatedly applying the recursion in Eq.~(\ref{nudrv}), we can express $\hat{\nu}_{m \in I_E}$ in terms of values at the base layer $n_{\rm e}$:
\begin{equation}
	\hat{\nu}_{m\in I_E}(\beta) \approx \frac{e^{-\alpha(E-n_{\rm e})}}{\binom{E}{n_{\rm e}}} \sum_{m'\in J^{n_{\rm e}}_E(m)} \hat{\nu}_{m'}(\beta), \quad \text{for } E > n_{\rm e}.
	\label{nubar_base}
\end{equation}
Here, $J^{n_{\rm e}}_E(m) = \{ m' \in I_{n_{\rm e}} \mid D_H(m', m) = E - n_{\rm e} \}$ is the set of states with energy $n_{\rm e}$ that has a Hamming distance $E-n_{\rm e}$ between state $m$. The size of this set is $|J^{n_{\rm e}}_E(m)| = \binom{E}{n_{\rm e}}$. Eq.~(\ref{nubar_base}) suggests that $\hat{\nu}_m$ for $m \in I_E$ can be viewed as an average over a specific subset of states in $I_{n_{\rm e}}$. Using concepts from finite population sampling theory \cite{Cochran1977} to relate the variance of such sample means to the population variance (within $I_{n_{\rm e}}$), we propose the following formula for the average second moment for $E > n_{\rm e}$:
\begin{equation}
	\bar{\mu}_E(\beta) \approx e^{-2\alpha(E-n_{\rm e})}\left[
	{\bar{\mu}_{n_{\rm e}}(\beta)}\,\frac{\binom{N}{n_{\rm e}}-\binom{E}{n_{\rm e}}}{\binom{E}{n_{\rm e}}\left(\binom{N}{n_{\rm e}}-1\right)}
	+\bar{\nu}^2_{n_{\rm e}}(\beta)\,\frac{\binom{N}{n_{\rm e}}\left(\binom{E}{n_{\rm e}}-1\right)}{\binom{E}{n_{\rm e}}\left(\binom{N}{n_{\rm e}}-1\right)}
	\right], \quad \text{for } E > n_{\rm e}.
	\label{mubar}
\end{equation}
This formula assumes $n_{\rm e} \ge 1$ and $\binom{N}{n_{\rm e}} > 1$. It relates $\bar{\mu}_E$ to the average $\bar{\nu}_{n_{\rm e}}(\beta)$ and the average second moment $\bar{\mu}_{n_{\rm e}}(\beta)$ at the base energy layer $n_{\rm e}$, which are computed directly from the presampling data:
\begin{equation}
	\bar{\nu}_{n_{\rm e}}(\beta) = \binom{N}{n_{\rm e}}^{-1} \sum_{m \in I_{n_{\rm e}}} \hat{\nu}_m(\beta), \quad
	\bar{\mu}_{n_{\rm e}}(\beta) = \binom{N}{n_{\rm e}}^{-1} \sum_{m \in I_{n_{\rm e}}} \hat{\mu}_m(\beta).
	\label{eq:base_averages}
\end{equation}
Equations (\ref{nubar_rec}) and (\ref{mubar}) allow us to extrapolate the necessary average quantities $\bar{\nu}_E(\beta)$ and $\bar{\mu}_E(\beta)$ for all energy levels $E > n_{\rm e}$ from the results obtained at the base layer $E=n_{\rm e}$ during presampling.

\textbf{4. Determination of Sampling Function:}
The parameter $\alpha$ required for the extrapolations, Eqs.~(\ref{nudrv})--(\ref{mubar}), is determined self-consistently. We use the analogue of the minimum variance condition Eq.~(\ref{alpeq2}), but replace the exact $\mu_m(\beta)$ with the estimated or extrapolated average second moments $\bar{\mu}_E(\beta)$:
\begin{equation}
	\frac{N}{1 + e^\alpha} = \sum_{E=0}^N E \cdot \bar{Q}_E(\alpha,\beta), \quad \text{where } \bar{Q}_E(\alpha,\beta) = \frac{\binom{N}{E} e^{\alpha E} \bar{\mu}_E(\beta)}{\sum_{E'=0}^N \binom{N}{E'} e^{\alpha E'} \bar{\mu}_{E'}(\beta)}.
	\label{eq:alpha_self_consistent_est}
\end{equation}
In this equation, $\bar{\mu}_E(\beta)$ for $E \le n_{\rm e}$ are calculated directly from the presampling data for $E\leq n_{\rm e}$, and for $E > n_{\rm e}$, they are obtained using the extrapolation formula Eq.~(\ref{mubar}). The parameter $\alpha$ is then determined by numerically solving Eq.~(\ref{eq:alpha_self_consistent_est}).

Once $\alpha$ is determined, we can compute $\hat{\nu}_m(\beta)$ values needed for sampling. For $m$ with $E_m^0 \le n_{\rm e}$, $\hat{\nu}_m(\beta)$ is obtained directly from presampling [Eq.~(\ref{eq:mu_hat})]. For $m$ with $E_m^0 > n_{\rm e}$, $\hat{\nu}_m(\beta)$ could, in principle, be calculated recursively using Eq.~(\ref{nudrv}) starting from the values at $E=n_{\rm e}$, although this may not be necessary for all states if their sampling probability is negligible.
The final sampling probability distribution $\mathcal{P}^{n_{\rm e}}_m(\beta)$ used in the main protocol is then constructed using these estimated $\hat{\nu}_m(\beta)$ values:
\begin{equation}
	\mathcal{P}^{n_{\rm e}}_m(\beta) = \frac{\hat{\nu}_m(\beta)}{\sum_{E=0}^{n_{\rm e}} \sum_{m^\prime \in I_E} \hat{\nu}_{m^\prime}(\beta) + \sum_{E=n_{\rm e}+1}^{N} \binom{N}{E} \bar{\nu}_E(\beta)}.
	\label{eq:Pm_final}
\end{equation}

With the sampling function $\mathcal{P}^{n_{\rm e}}_m(\beta)$ defined by Eq.~(\ref{eq:Pm_final}), we estimate the partition function by sampling trajectories $|{\psi^0_m}\rangle\to |{\psi^1_n}\rangle$ with probability $P_{n|m} \mathcal{P}^{n_{\rm e}}_m(\beta)$, as illustrated in Eq.~(3) of the main text. We find that even for small integers $n_{\rm e}$ (e.g., $n_{\rm e}=1, 2$), the resulting sampling function $\mathcal{P}^{n_{\rm e}}_m(\beta)$ exhibits good performance, as shown in Fig.~3 of the main text.

\section{V. Complexity analysis}

In this section, we analyze the complexity of the partition function estimation protocol with respect to the Hilbert space dimension $D=2^N$.

When estimating the Ising partition function $Z_1(\beta)$ of $H_1$, we sample $M_{\mathrm{s}}$ trajectories $|{\psi^0_m}\rangle\to |{\psi^1_n}\rangle$, where $|{\psi^0_m}\rangle$ is chosen according to $P_m$ [ideally $\mathcal{P}_m(\beta)$ or our approximation $\mathcal{P}^{n_{\rm e}}_m(\beta)$] and $P_{n|m}$ is the Born probability of eigenbasis measurement of $H_{1}$. The partition function is estimated as
\begin{equation}
	Z_{\text {est }}(\beta) = \frac{1}{M_{\mathrm{s}}} \sum_{i=1}^{M_{\mathrm{s}}} z_{m_i, n_i}(\beta),
\end{equation}
where $z_{m,n}(\beta) = {e^{-\beta E^1_n}}/{P_m}$ and $(m_i, n_i)$ denotes the initial and final states obtained in the $i$-th sample run.
As different runs are independent, the variance of the estimator $Z_{\text{est}}(\beta)$ scales inversely with the sample size,
\begin{equation}
	\mathrm{Var}[Z_{\text {est }}(\beta)] = \frac{\sigma^2(\beta)}{M_{\mathrm{s}}},
\end{equation}
where $\sigma^2(\beta) = \mathrm{Var}[z_{m,n}(\beta)]$ is given by Eq.~(\ref{sigma2z}).
Requiring a prescribed relative standard error $\sqrt{\mathrm{Var}[Z_{\text {est }}(\beta)]} / Z_1(\beta) = \varepsilon$ implies
\begin{equation}
	M_{\mathrm{s}} = \frac{\sigma^2(\beta)}{\varepsilon^2 Z_1^2(\beta)}.
\end{equation}
Therefore, the relative variance $\sigma^2(\beta)/Z_1^2(\beta)$ characterizes the sampling complexity of the protocol (up to the cost per sample).

\textbf{Scaling of the minimum variance with $\mathcal{P}_m(\beta)$.}
We first analyze how the minimum achievable relative variance $\sigma^2_{\mathrm{min}}(\beta)/Z_1^2(\beta)$, given by Eq.~(\ref{varmin}), scales with the Hilbert space dimension $D = 2^N$. We assume a relationship of the form $\sigma^2_{\mathrm{min}}(\beta)/Z_1^2(\beta) \sim D^\gamma = 2^{N\gamma}$ for large $N$, where $\gamma$ is a scaling exponent. First, an upper bound can be derived using the Cauchy--Schwarz inequality. Applying it to the sum in Eq.~(\ref{varmin}), we find
\begin{equation}
	\sigma^2_{\mathrm{min}}(\beta)/Z_1^2(\beta)<\frac{1}{Z_1^2(\beta)}\left( \sum_{m=1}^D \sqrt{\mu_m(\beta)} \right)^2 \le \frac{1}{Z_1^2(\beta)}\left( \sum_{m=1}^D (\sqrt{\mu_m(\beta)})^2 \right) \left( \sum_{m=1}^D 1^2 \right) = \frac{Z_1(2\beta)}{Z_1^2(\beta)}D<\frac{D}{Z_1(\beta)}.
	\label{sg2neq}
\end{equation}

In the high-temperature limit ($\beta \to 0$), $Z_1(\beta) \to D$ and $Z_1(2\beta) \to D$. The bound becomes $D \cdot D / D^2 = 1$. Since $D^\gamma \sim 1$, this implies $\gamma \to 0$ as $\beta \to 0$, indicating low complexity at high temperatures. We are more interested in the low-temperature regime (large $\beta$), which is typically harder for estimation problems. In this regime, since the ground state degeneracy is at least $1$, the minimum achievable variance always yields $\gamma \le 1$. Moreover, it is also expected that $\sigma^2_{\mathrm{min}}(\beta)/Z_1^2(\beta)$ saturates for large $\beta$ from Eq.~(\ref{sg2neq}), which can also be inferred from the saturation of $\alpha$ in the Gibbs distribution approximation (Fig.~\ref{figS2}).

To get a better estimate of $\gamma$, we draw from the observation that the optimal sampling probability $\mathcal{P}_m(\beta) \propto \sqrt{\mu_m(\beta)}$ decays approximately exponentially with initial energy $E_m^0$ at a rate of $\alpha$, i.e., $\sqrt{\mu_m(\beta)} \approx \lambda \exp(-\alpha E_m^0)$, where $\lambda$ is a normalization constant. To determine $\lambda$, we use the approximate relationship $\sum_m \mu_m(\beta) \approx Z_1(2\beta)$.
\begin{equation}
	\sum_{m=1}^D \mu_m(\beta) = \sum_{m=1}^D (\sqrt{\mu_m(\beta)})^2 \approx \sum_{m=1}^D \lambda^2 e^{-2\alpha E_m^0} = \lambda^2 Z_0(2\alpha) = \lambda^2 (1 + e^{-2\alpha})^N.
\end{equation}
Setting this equal to $Z_1(2\beta)$, we get $\lambda^2 \approx Z_1(2\beta) / (1+e^{-2\alpha})^N$.
Now we estimate the dominant term in the variance, $(\sum \sqrt{\mu_m})^2$:
\begin{equation}
	(\sum_{m=1}^D \sqrt{\mu_m(\beta)})^2 \approx (\sum_{m=1}^D \lambda e^{-\alpha E_m^0})^2 = \lambda^2 (Z_0(\alpha))^2
	\approx \frac{Z_1(2\beta)}{(1+e^{-2\alpha})^N} (1+e^{-\alpha})^{2N} = Z_1(2\beta) \left( \frac{(1+e^{-\alpha})^2}{1+e^{-2\alpha}} \right)^N.
\end{equation}
Assuming $\sigma^2_{\mathrm{min}}(\beta) \approx (\sum \sqrt{\mu_m})^2$, the relative variance scales as:
\begin{equation}
	\sigma^2_{\mathrm{min}}(\beta)/Z_1^2(\beta) \approx \frac{Z_1(2\beta)}{Z_1^2(\beta)} \left( \frac{(1+e^{-\alpha})^2}{1+e^{-2\alpha}} \right)^N = \frac{Z_1(2\beta)}{Z_1^2(\beta)} \left( 1 + \frac{2e^{-\alpha}}{1 + e^{-2\alpha}} \right)^N.
	\label{sg2apr}
\end{equation}
At the low temperature case, $Z_1(2\beta)/Z_1^2(\beta)$ is sub-exponential in $N$ , the dominant scaling comes from the exponential term. Comparing $D^\gamma = 2^{N\gamma}$ with the base of the exponential term in Eq.~(\ref{sg2apr}), we get:
\begin{equation}
	\gamma = \log_2 \left( 1 + \frac{2e^{-\alpha}}{1 + e^{-2\alpha}} \right).
	\label{gamapr}
\end{equation}
In our numerical results for $N=8$ (at $\beta=10$, averaged over 1000 instances), we found $\alpha \approx 1.921$ for the SK model and $\alpha \approx 1.575$ for the 3-SAT model (using minimum variance criterion, red circles in Fig.~\ref{figS2}). Plugging these values into Eq.~(\ref{gamapr}), we arrive at the following scaling exponents:
\begin{itemize}
	\item SK model: $\gamma \approx \log_2(1 + \frac{2e^{-1.921}}{1+e^{-3.842}}) \approx \log_2(1.287) \approx 0.364$;
	\item 3-SAT model: $\gamma \approx \log_2(1 + \frac{2e^{-1.575}}{1+e^{-3.150}}) \approx \log_2(1.397) \approx 0.482$.
\end{itemize}
These theoretical estimates ($\gamma \approx 0.364$ for SK, $\gamma \approx 0.482$ for 3-SAT) agree well with the numerically fitted scaling exponents derived from the optimal sampling function $\mathcal{P}_m(\beta)$, as shown by the red solid lines in Figs.~2(a2) and 2(b2) of the main text ($\gamma \approx 0.354$ for SK and $\gamma \approx 0.473$ for 3-SAT).

\textbf{The effect of finite presampling depth $n_{\rm e}$.}
For the practical sampling functions $\mathcal{P}^{n_{\rm e}}_m$ generated with a finite presampling depth $n_{\rm e}$,
the parameter $\alpha$ is inferred self-consistently from presampling via Eq.~(\ref{eq:alpha_self_consistent_est}).
As a result, the estimated $\alpha'$ generally deviates from the ideal $\alpha$ obtained from Eq.~(\ref{alpeq2}),
which increases the effective complexity.
To quantify how $\alpha'$ impacts the variance, we use the same exponential approximation
$\sqrt{\mu_m(\beta)}\approx \lambda e^{-\alpha E_m^0}$ with
$\lambda^2 \approx Z_1(2\beta)/(1+e^{-2\alpha})^N$,
and approximate the implemented sampling distribution by
$P_m \approx e^{-\alpha' E_m^0}/Z_0(\alpha')$.
Using the bound $\sigma^2(\beta)/Z_1^2(\beta)<Z_1(\beta)^{-2}\sum_m \mu_m(\beta)/P_m$, we obtain
\begin{equation}
	\frac{\sigma^2(\beta)}{Z_1^2(\beta)}
	<\frac{1}{Z_1^2(\beta)}\sum_{m=1}^D \frac{\mu_m(\beta)}{P_m}
	\approx
	\frac{Z_1(2\beta)}{Z_1^2(\beta)}
	\left(\frac{(1+e^{-\alpha'})(1+e^{-(2\alpha-\alpha')})}{1+e^{-2\alpha}}\right)^N
	=
	\frac{Z_1(2\beta)}{Z_1^2(\beta)}
	\left(1+\frac{e^{-\alpha'}+e^{-(2\alpha-\alpha')}}{1+e^{-2\alpha}}\right)^N.
	\label{eq:finite_depth_bound}
\end{equation}
This motivates the definition of an effective exponent $\gamma'$
\begin{equation}
	\gamma^{\prime} = \log_2\!\left(1+\frac{e^{-\alpha'}+e^{-(2\alpha-\alpha')}}{1+e^{-2\alpha}}\right),
	\label{eq:gamma_prime}
\end{equation}
which satisfies $\gamma'>\gamma$ whenever $\alpha'\neq \alpha$, and approaches $\gamma$ as $\alpha'\to \alpha$.
For $\mathcal{P}^{n_{\rm e}}_m$ with $n_{\rm e}=1$, our numerics for $N=8$ at $\beta=10$ (averaged over $1000$ instances) give
$\gamma^\prime\approx 0.422$ for SK and $\gamma^\prime\approx 0.521$ for 3-SAT,
comparable to the exponents measured directly from the practical protocol with $n_{\rm e}=1$
($\gamma' \approx 0.446$ for SK and $\gamma' \approx 0.502$ for 3-SAT).
This indicates that, even with minimal presampling, the protocol achieves scaling close to the optimal one in our numerics.
As $n_{\rm e}$ increases, $\mathcal{P}^{n_{\rm e}}_m$ approaches the optimal $\mathcal{P}_m$,
and the inferred $\alpha'$ becomes closer to $\alpha$, thereby recovering the ideal sampling complexity.

Finally, Eqs.~(\ref{gamapr}) and (\ref{eq:gamma_prime}) show that $\gamma$ decreases with increasing $\alpha$ (and, in the finite-$n_{\rm e}$ case, $\gamma'$ follows the same trend since $\alpha'\approx\alpha$).
As observed in Fig.~\ref{figS2} [and Fig.~2(b) in the main text], a longer annealing time $\tau$ leads to better adiabaticity (less excitations to higher energy states of $H_1$), which typically results in a larger effective $\alpha$ (steeper decay of $\sqrt{\mu_m}$ with $E_m^0$). This larger $\alpha$ leads to a lower complexity exponent $\gamma$, which suggests that achieving optimal performance for larger system sizes may require longer annealing times $\tau$ and thus better coherence.

Therefore, for a given accuracy $\varepsilon$ such that
$\frac{|Z_{\mathrm{est}}(\beta)-Z_1(\beta)|}{Z_1(\beta)} < \varepsilon,$
the computational complexity of our algorithm scales as
\begin{equation}
	M_s \sim \mathcal{O}\!\left(\frac{D^\gamma}{\varepsilon^2 Z_1(\beta)}\right),
\end{equation}
where the exponent $\gamma$ is determined by the structure of the Hamiltonian and is strictly bounded by $\gamma < 1$. For paradigmatic hard models such as the Sherrington–Kirkpatrick (SK) spin glass and random 3-SAT Hamiltonians, we observe that $\gamma \leq 0.5$ even with very shallow presampling depth $n_{\rm e}$, indicating that our approach substantially reduces the effective exponential scaling.

\section{VI. Complexity comparison with other algorithms}
In this section, we compare our algorithm's complexity against existing sampling-based methods for partition function estimation, focusing on the challenging low-temperature regime.

Estimating the partition function of a general Hamiltonian is a \#P-hard problem~\cite{Lidar_2004_NJP}, for which no efficient classical algorithm is believed to exist. Classical approaches typically rely on Markov Chain Monte Carlo (MCMC) simulated annealing~\cite{Bezkov2008,tefankovi2009,kolmogorov18a}, where evaluation of the partition function $Z(\beta_\ell)$ is obtained by implementing a cooling sequence $0=\beta_0<\beta_1<\dots<\beta_\ell=\beta$ and successively estimating the partition function at lower temperatures. With this divide-and-conquer approach, the partition function is expressed as
$Z(\beta_\ell) = Z(\beta_0) \prod_{i=1}^\ell \frac{Z(\beta_i)}{Z(\beta_{i-1})}$.
The optimal complexity of known classical MCMC algorithms scales as $\tilde{\mathcal{O}}(\tfrac{\log |D|}{\varepsilon^2}\tau)$~\cite{kolmogorov18a}, where $\tau$ denotes the Markov chain mixing time. While $\tau$ may scale polynomially in simple cases, it is exponentially large in frustrated or glassy systems such as SK or random SAT: The asymptotic behavior of mixing time reads $\tau \sim e^{c(\beta) N}$ with $c(\beta)$ typical linearly growing with $\beta$ at low temperatures \cite{BenArous2018,Gheissari2019}. This exponential mixing-time barrier renders MCMC-based approaches impractical for low-temperature partition function estimations.

Beyond standard simulated annealing, two widely used classical families for navigating complex energy landscapes are annealed importance sampling (AIS)~\cite{Neal2001AIS} and generalized-ensemble (flat-histogram) methods. The latter includes well-established techniques such as multicanonical sampling~\cite{BergNeuhaus1992} and the Wang--Landau algorithm~\cite{WangLandau2001a,WangLandau2001b}. AIS interleaves MCMC transitions along a temperature ladder with importance reweighting. In a single AIS run, an importance weight is accumulated as $w=\prod_{i=1}^{\ell}\exp\!\big[-(\beta_i-\beta_{i-1})\,E(x_{i-1})\big]$, where $x_{i-1}$ is obtained by applying a chosen MCMC transition kernel targeting (or approximating) the Gibbs distribution at inverse temperature $\beta_{i-1}$. The partition function is then estimated via $Z(\beta)=Z(\beta_0)\,\mathbb{E}[w]$ by averaging over independent AIS runs. A common cost model for achieving relative error $\varepsilon$ takes the form $\tilde{\mathcal{O}}\!\left(\frac{1}{\varepsilon^2}\sum_{i=1}^{\ell}\tau_{\rm mix}(\beta_i)\right)\equiv\tilde{\mathcal{O}}\!\left(\frac{\ell}{\varepsilon^2}\tau_{\rm eff}\right)$, where $\tau_{\rm mix}(\beta_i)$ denotes the effective equilibration/mixing (or relaxation) effort required at $\beta_i$ to obtain near-equilibrium samples with sufficiently controlled correlations, and $\tau_{\rm eff}$ is the effective average along the ladder. In rugged, low-temperature landscapes, reliable performance typically requires both (i) adequate overlap between successive distributions (often favoring a longer ladder, large $\ell$) and (ii) sufficient relaxation at each stage; consequently, $\tau_{\rm eff}$ can inherit the same barrier-crossing difficulties that impede standard annealing.

Generalized-ensemble methods adopt a complementary strategy by estimating the density of states $g(E)$ (up to normalization), from which the partition function is reconstructed as $Z(\beta)=\sum_E g(E)\,e^{-\beta E}$. By iteratively refining weights to produce an approximately flat energy histogram, multicanonical sampling and the Wang--Landau algorithm can facilitate repeated excursions across the relevant energy range and thereby mitigate trapping relative to naive annealing. A convenient performance indicator is the round-trip time $\tau_{\rm rt}$ of the induced random walk to traverse the relevant energy range. However, even under idealized assumptions with accurate weights, $\tau_{\rm rt}$ can grow rapidly and exhibit broad instance-to-instance fluctuations in disordered systems such as spin glasses~\cite{Dayal2004WLspinGlass}. While feedback-optimized variants can improve sampling efficiency~\cite{Trebst2004OptimizedEnsembles,Katzgraber2006WL}, the overall cost remains governed by extensive exploration of the energy space, which becomes increasingly demanding at low temperatures.

Consequently, although AIS and generalized-ensemble methods often improve performance over naive annealing, they do not generally eliminate the low-temperature exploration bottleneck in frustrated/glassy systems.

Quantum algorithms based on quantum walks and amplitude estimation~\cite{Wocjan2009,Montanaro2015,Harrow2020,Arunachalam2022,Cornelissen2023} can quadratically accelerate MCMC sampling, with complexity scaling as $\tilde{\mathcal{O}}(\frac{\log^{3/4}|D|}{\varepsilon}\sqrt{\tau})$. However, since $\tau$ remains exponential for glassy systems, the overall scaling is still dominated by $e^{c(\beta)N/2}$, meaning the exponential barrier persists.

Another line of quantum approaches exploits quantum phase estimation (QPE)~\cite{Poulin2009}. For Ising-type partition functions, the complexity scales as $\tilde{\mathcal{O}}(\tfrac{\beta}{\varepsilon}\sqrt{\frac{D}{Z_1(\beta)}})$, yielding a quadratic improvement in sampling accuracy. This advantage, however, requires extremely deep quantum circuits, high-precision controlled unitaries, and long coherent evolution times, each of which poses significant experimental challenges.

Alternative strategies based on deterministic quantum computation with one clean qubit (DQC1)~\cite{Chowdhury2021,Jackson2023} estimate the partition function through trace estimation of the evolution operator. Their complexity scales as $\tilde{\mathcal{O}}(\tfrac{D^2\beta^2}{\varepsilon^2Z_1(\beta)^2})$, which is substantially worse than ours. Also, these algorithms require a large-scale maximally mixed register and high-fidelity controlled operations, posing severe experimental obstacles. An improved scheme using a quantum coin~\cite{Silva2411} reduces the scaling to $\tilde{\mathcal{O}}(\tfrac{D e^\beta}{\varepsilon^2Z_1(\beta)})$, but still remains more costly than our approach.

In contrast, our algorithm avoids the long mixing-time bottleneck intrinsic to Markov-chain exploration (including sophisticated variants such as AIS and flat-histogram generalized ensembles), requires neither QPE nor deep amplitude estimation, and achieves a favorable complexity scaling. The effective exponent $\gamma < 1$ guarantees sub-exponential dependence on the Hilbert space dimension $D$, and for practically relevant hard models we find $\gamma \leq 0.5$. This places our algorithm in a distinct complexity class, offering a significantly more favorable trade-off between accuracy and computational resources in the low-temperature regime where existing algorithms face exponential slowdowns. A summary of these complexity comparisons is provided in Table~\ref{tab:complexity}.

\renewcommand{\arraystretch}{1.65}
\setlength{\tabcolsep}{8pt}
\begin{table}[h]
	\centering
	\caption{Comparison of computational complexities for partition function estimation algorithms.}
	\label{tab:complexity}
	\begin{tabular}{p{5.7cm}| p{2.5cm}| p{6.8cm}}
		\hline
		Method \& References & Complexity & Comment \\
		\hline
		\raisebox{-1.1em}{Classical MCMC annealing~\cite{Bezkov2008,tefankovi2009,kolmogorov18a}}
		& \raisebox{-1.1em}{$\tilde{\mathcal{O}}\!\left(\tfrac{\log |D|}{\varepsilon^2}\tau\right)$}
		& Exponentially large mixing time at low $T$ for frustrated/glassy systems (e.g., SK, random SAT). \\
		\hline
		\raisebox{-0.9em}{AIS (annealed importance sampling)~\cite{Neal2001AIS}}
		& \raisebox{-0.9em}{$\tilde{\mathcal{O}}\!\left(\tfrac{\ell}{\varepsilon^2}\,\tau_{\mathrm{eff}}\right)$}
		& Requires adequate equilibration across a temperature ladder; $\tau_{\mathrm{eff}}$ inherits low-$T$ barrier issues in rugged landscapes. \\
		\hline
		\raisebox{-1.0em}{Multicanonical / Wang--Landau~\cite{BergNeuhaus1992,WangLandau2001a,WangLandau2001b}}
		& \raisebox{-1.0em}{$\tilde{\mathcal{O}}\!\left(\tfrac{1}{\varepsilon^2}\,\tau_{\mathrm{rt}}\right)$}
		& Density-of-states learning governed by tunneling/round-trip time; can be strongly slowed in glassy landscapes~\cite{Dayal2004WLspinGlass}; optimized variants still require extensive exploration~\cite{Trebst2004OptimizedEnsembles,Katzgraber2006WL}. \\
		\hline
		Quantum MCMC (quantum walks + amplitude estimation)~\cite{Wocjan2009,Montanaro2015,Harrow2020,Arunachalam2022,Cornelissen2023}
		& \raisebox{-0.5em}{$\tilde{\mathcal{O}}\!\left(\tfrac{\log^{3/4}|D|}{\varepsilon}\sqrt{\tau}\right)$}
		& Quadratic speedup in $\tau$, but $\tau$ still diverges exponentially for glassy systems. \\
		\hline
		\raisebox{-0.5em}{Quantum phase estimation (QPE)~\cite{Poulin2009}}
		& \raisebox{-0.5em}{$\tilde{\mathcal{O}}\!\left(\tfrac{\beta}{\varepsilon}\sqrt{\tfrac{D}{Z_1(\beta)}}\right)$}
		& Quadratic accuracy gain; requires deep circuits and precise controlled unitaries. \\
		\hline
		\raisebox{-0.5em}{DQC1-based algorithms~\cite{Chowdhury2021,Jackson2023}}
		& \raisebox{-0.5em}{$\tilde{\mathcal{O}}\!\left(\tfrac{D^2\beta^2}{\varepsilon^2 Z_1(\beta)^2}\right)$}
		& Needs large maximally mixed register and high-fidelity gates; scaling worse than ours. \\
		\hline
		\raisebox{-.3em}{Quantum coin scheme~\cite{Silva2411}}
		& \raisebox{-.3em}{$\tilde{\mathcal{O}}\!\left(\tfrac{D e^\beta}{\varepsilon^2 Z_1(\beta)}\right)$}
		& \raisebox{-.3em}{Improves over DQC1 but still costly.} \\[6pt]
		\hline
		\raisebox{-1.0em}{Our algorithm}
		& \raisebox{-1.0em}{$\mathcal{O}\!\left(\tfrac{D^\gamma}{\varepsilon^2 Z_1(\beta)}\right)$ }
		& Avoids Markov-chain exploration bottlenecks; no QPE or deep circuits; sub-exponential in $D$, with $\gamma \leq 0.5$ for SK and 3-SAT. \\
		\hline
	\end{tabular}
\end{table}

\bibliography{reference.bib}